\documentclass[journal,12pt,draftclsnofoot,onecolumn]{IEEEtran}

\usepackage{cite}

\usepackage{makeidx,epsfig}
\usepackage{setspace,graphicx,multirow}%,subfigure

\usepackage{amsfonts,latexsym,amssymb,amsthm}

\usepackage{graphicx}

\usepackage{caption3} % load caption package kernel first
\DeclareCaptionOption{parskip}[]{} % disable "parskip" caption option
\usepackage[small]{caption}

\usepackage{subcaption}

\usepackage{array}

\usepackage[cmex10]{amsmath}

\usepackage{url}

\usepackage{multirow}% http://ctan.org/pkg/multirow
\usepackage{hhline}% http://ctan.org/pkg/hhline

\usepackage{amsmath}
\usepackage{algorithm}
\usepackage[]{algpseudocode}
\usepackage{varwidth}

\makeatletter
\def\BState{\State\hskip-\ALG@thistlm}
\makeatother

\begin{document}
\bibliographystyle{IEEEtran}
\bstctlcite{IEEEexample:BSTcontrol}

\title{A Stackelberg Game Model for Overlay D2D Transmission with Heterogeneous Rate Requirements}

\author{Jiangbin~Lyu,
        Yong~Huat~Chew,~\IEEEmembership{Member,~IEEE}
        and~Wai-Choong~Wong,~\IEEEmembership{Senior~Member,~IEEE}%
%\thanks{Manuscript received March 24, 2015; revised August 19 and October 14, 2015; accepted December 11, 2015. The editor coordinating the review of this paper and approving it for publication was Prof. Yi Qian.}%
\thanks{J. Lyu is with NUS Graduate School for Integrative Sciences and Engineering (NGS), National University of Singapore (email: jiangbin.lu@u.nus.edu)}%
\thanks{Y. H. Chew is with the Institute for Infocomm Research, Singapore (email: chewyh@i2r.a-star.edu.sg)}%
\thanks{W. C. Wong is with the Department of Electrical and Computer Engineering, National University of Singapore (email: elewwcl@nus.edu.sg)}
%\thanks{Digital Object Identifier 10.1109/TVT.2015.000000.000000} 
}

%\markboth{IEEE Transactions on Vehicular Technology,~Vol.~XX, No.~X, XXXXX~XXXX}%
%{Lyu \MakeLowercase{\textit{et al.}}: A Stackelberg Game Model for Overlay D2D Transmission with Heterogeneous Rate Requirements}

%\IEEEpubid{
%\begin{varwidth}[t]{2.1\linewidth}
%\centering\copyright~2015 IEEE. Personal use of this material is permitted. \\
%However, permission to use this material for any other purposes must be obtained from the IEEE by sending a request to pubs-permissions@ieee.org.
%      \end{varwidth}
%}


\maketitle

%\vspace{-5pt}
\begin{abstract}
This paper studies the performance of overlay device-to-device (D2D) communication links via carrier sense multiple access (CSMA) protocols. We assume that the D2D links have heterogeneous rate requirements and different willingness to pay, and each of them acts non-altruistically to achieve its target rate while maximizing its own payoff. Spatial reuse is allowed if the links are not interfering with each other. A non-cooperative game model is used to address the resource allocation among the D2D links, at the same time leveraging on the ideal CSMA network (ICN) model to address the physical channel access issue. We propose a Stackelberg game in which the base station in the cellular network acts as a Stackelberg leader to regulate the individual payoff by modifying the unit service price so that the total D2D throughput is maximized.
The problem is shown to be quasi-convex and can be solved by a sequence of equivalent convex optimization problems. The pricing strategies are designed so that the network always operates within the feasible throughput region. The results are verified by simulations.
\end{abstract}


\section{Introduction}

The growing popularity of smartphones and tablets has resulted in the increasing demand for high data rate services, and a huge amount of data traffic normally needs to be transmitted through cellular networks, which in turn leads to severe traffic overload problems. Recently, Device-to-Device (D2D) communication has emerged as a new data-offloading solution by enabling direct communication between two mobile users without traversing the base station (BS) \cite{D2Dsurvey}.

D2D communication can be implemented over the cellular spectrum (i.e. inband) or the unlicensed spectrum (i.e. outband). 
Inband D2D can be further classified into spectrum overlay and spectrum underlay. 
In the overlay scenario, the cellular and D2D transmitters use orthogonal time/frequency resources, while in the underlay scenario the D2D transmitters access the same time/frequency resources occupied by cellular users.
The rate performance is evaluated in \cite{D2DoverlayAndrews} for both overlay and underlay scenarios. It is observed that D2D mobiles in both scenarios can enjoy much higher data rates than regular cellular mobiles. As for cellular mobiles in the overlay scenario, their rate performance also improves due to the offloading capability of D2D communication.
Besides performance improvement over the pure cellular mode, inband overlay D2D is also more tractable in analysis since it does not interfere with regular cellular mobiles or suffer from random interference from unlicensed band.

\begin{figure}[t]
\centering
\includegraphics[width=0.7\linewidth,  trim=0 90 0 0,clip]{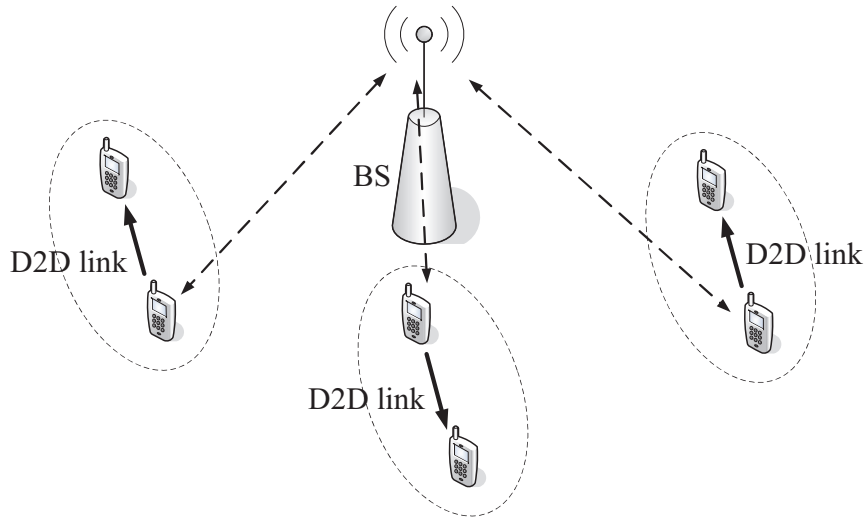} %,height=0.5\linewidth
\caption{An Example of Overlay D2D Communications. The solid arrow represents the data communication of a D2D link; the dashed arrow represents the control information exchange between a D2D transmitter and the BS.} \label{D2D}
\vspace{-1em}
\end{figure}

In \cite{D2DoverlayAndrews} the authors use a simple spatial Aloha access scheme to support D2D scheduling.
In this paper we assume that all D2D links use carrier sense multiple access (CSMA) as the multiple access scheme to share a dedicated inband overlay channel.   
Spatial reuse is considered, i.e., different transmit-receive (Tx-Rx) pairs at a sufficient distance away that do not cause interference are allowed to transmit simultaneously\cite{LiewBoE}. Although D2D communication does not route the data traffic through the cellular network, the available network infrastructure can still be an effective means to exert light control over all the D2D links when performing resource allocation. In our model, the D2D links have heterogeneous service requirements and different willingness to pay, and the central entity (e.g., evolved node B (eNB))\cite{D2Dsurvey} controls the transmission behaviors of all links by modifying the price per unit service rate. 
A simple example is given in Fig. \ref{D2D}, where there are three D2D links and a single BS oversees/controls them in the control plane. Each D2D link consists of a Tx-Rx pair, and hence the D2D links resemble the situation where distributed pairs are transmitting. Hereafter, the terms ``D2D link", ``CSMA Tx-Rx pair" and ``CSMA user" are used synonymously. The involvement of the cellular network in the control plane is the key difference between our system model with that defined in Mobile Ad-hoc NETworks (MANET)\cite{MANET}. Moreover, D2D communication is mainly used for single-hop communications which does not inherit the multihop routing problem of MANET and wireless sensor networks\cite{JMchenDataGatheringTNET}.% \cite{MANET}

\IEEEpubidadjcol

When spatial reuse is considered in CSMA methods, the carrier-sense relationships among the CSMA users become \textit{non-all-inclusive}, i.e., each CSMA user may only sense a subset, but not all other users. As commented by Liew \textit{et al.}\cite{LiewBoE}, it is extremely difficult to extend the analytic methods for all-inclusive carrier-sense networks (e.g., \cite{Bianchi,YangXiaoDCFtvt,BianchiDCFrefineTVT}) to the non-all-inclusive case because of the inhomogeneity in the state spaces of the CSMA users. In fact, the problem of computing user throughputs in a spatial CSMA network is shown to be NP-hard\cite{LiewBoE}. In order to perform tractable analysis, existing literature \cite{FirstIdealCSMA,KarCSMA, LiewBoE, JiangLibinCSMA} have adopted an \textit{ideal CSMA network (ICN)} model to capture the essence of the CSMA mechanism under spatial reuse. %FirstIdealCSMA,
In this paper we also leverage on the ICN model to address the physical channel access issue.

%%%%Need more expansion below to present the MAIN TRICK to attract people, i.e., make them feel good: I can understand the main trick!!! I want to know more details!!!
Our contributions in this paper are twofold. First of all, we propose a Stackelberg game\cite{Fudenberg} which maximizes the total throughput of the D2D links, where these links have heterogeneous utility functions. The BS in the cellular networks will act as a Stackelberg leader to regulate the D2D link transmissions by modifying the service price, so that the payoff of each individual D2D link can be maximized while maintaining the D2D network to function within the feasible throughput region determined by the CSMA access mechanism. The problem is shown to be quasi-convex and can be solved by a sequence of equivalent convex optimization problems. The pricing strategies are designed so that the network always operates within the feasible throughput region. 
Secondly, each D2D link will acquire a rate based on its actual demand and willingness to pay. % and the equilibrium solution provides good fairness and differentiated services for the heterogeneous users. 
We explicitly model the possible selfish behaviors among the D2D links with spatial reuse. Under a given network price, the transmitter of each D2D link competes for channel usage by choosing its transmission parameters in order to maximize its own payoff. Such user dynamics are studied in the setting of non-cooperative games, and the resulting CSMA game model serves as the follower-subgame in the proposed Stackelberg game. An algorithm is proposed followed by proofs for the existence and convergence of the equilibrium solution.

The rest of the paper is organized as follows. We introduce related works in Section \ref{RelatedWorks} and the network model in Section \ref{CSMAModel}, and summarize some important results on ICN. In Section \ref{SectionFeasibleThroughput}, the feasible throughput region for a CSMA network is defined, while some important properties are derived. The Stackelberg game is detailed in Section \ref{Stackelberg}. Performance of the proposed game is evaluated through simulations in Section \ref{Simulation}. We conclude the paper in Section \ref{Summary}.

The notations used in this paper are as follows. An arrow over a variable $\vec{\cdot}$ represents a vector, or a system state consisting of the binary status of the $N$ links. The variable $\theta$ is used to denote the throughput from the channel access point of view, where the solution is controlled by the ICN model. On the other hand, $\tilde{\theta}$ is used to represent a desired throughput from the link layer aspect, whose value is derived from the price and link utility function. If the final solution is within the feasible throughput region, these two values should match. There are two types of equilibria to be differentiated: the subgame equilibrium is denoted by a superscript `*' whilst for the Stackelberg game is denoted by a superscript `opt'.  

\section{Related Works}\label{RelatedWorks}

\subsection{Spectrum Sharing Games for D2D Communication}
In the licensed spectrum, a potential D2D pair can communicate through conventional cellular mode (relay through the BS), dedicated D2D mode (spectrum overlay), or underlay sharing mode (share with cellular users).
Game theoretic approaches have been applied in D2D communications for mode selection and resource management \cite{GameD2Doverview}. 
%The potential D2D users can operate in cellular mode (relay through the BS), dedicated D2D mode (spectrum overlay), or sharing mode (spectrum underlay).
In particular, Cai \textit{et al}. \cite{CaiYuemingICC} model the spectrum sharing mode selection as a coalition formation game, and propose a distributed coalition formation algorithm to improve the total achievable rate. Wu \textit{et al}. \cite{WuDanTVT} study the underlay spectrum sharing problem among potential D2D pairs and cellular users with quality-of-service requirements. A coalition formation game and a distributed coalition formation algorithm are proposed to decide for the most energy-efficient spectrum sharing strategy.
The focus of these works is to look for efficient spectrum sharing solutions among the D2D pairs and cellular users, and spectrum underlay is adopted in the sharing mode under the constraints on the amount of mutual interference. In this paper, we focus on spectrum overlay mode, in which the number of orthogonal channels is limited and multiple D2D pairs share a common channel via distributed transmission scheduling.

\subsection{CSMA Distributed Transmission Scheduling}

A survey on applying game theory to CSMA can be found in \cite{GameCSMAsurvey}, where several non-cooperative contention control games in CSMA methods are presented. For example, Jin and Kesidis \cite{JinKesidisCSMA} analyze the non-cooperative user behaviors in CSMA wireless networks where users have the freedom to choose the contention window sizes according to the network congestion level. The existence and uniqueness of the equilibrium point are investigated, as well as a distributed iterative method to approach the equilibrium. However, as commented by \cite{GameCSMAsurvey}, most of the proposed CSMA games assume all-inclusive carrier sensing.
The analysis cannot be directly applied in the presence of spatial reuse.

The ICN model captures the essence of spatial reuse CSMA networks. Jiang and Walrand \cite{JiangLibinCSMA} developed an elegant distributed CSMA algorithm for throughput and total utility maximization based on the ICN model after making assumptions about concavity and monotonicity of the user utility functions. Their work removes the need for knowledge of the underlying link topology and their transmission parameters can be updated distributively. However, the approach implicitly assumes a best effort transmission to achieve total utility maximization and there is no explicit treatment if users have heterogeneous rate requirements and different willingness to pay. In other words, while optimizing the sum-rate, there is no mechanism to weigh the individual user utility %based on the his demand and willingness to pay for the bandwidth 
so as to differentiate the services. Moreover, the global optimization approach does not reflect the fact that users are selfish and behave non-altruistically in maximizing their own payoffs. In fact, Cagalj \textit{et al.} have shown that even the presence of a few selfish users may lead such a CSMA network to collapse\cite{SelfishCSMA}, while proper pricing or penalty mechanisms lead to overall improvement \cite{PenaltyWLAN}.
Indeed, when users with heterogeneous rate requirements coexist in the network and the collective target rates are outside the feasible throughput region, the self-interested actions by the CSMA users would lead the network into heavily congested status.
%We demonstrate this undesirable network behavior via an example in Section \ref{oscilate}.
In this paper we incorporate a game theoretic framework into the ICN model to harness the selfish behaviors of a group of non-cooperative spatially distributed CSMA users with heterogeneous rate requirements.

\section{Spatial Reuse CSMA Network}\label{CSMAModel}

\subsection{Spatial Reuse and Contention Graph}

\begin{figure}[t]
\centering
\includegraphics[width=0.55\linewidth,  trim=0 0 0 0,clip]{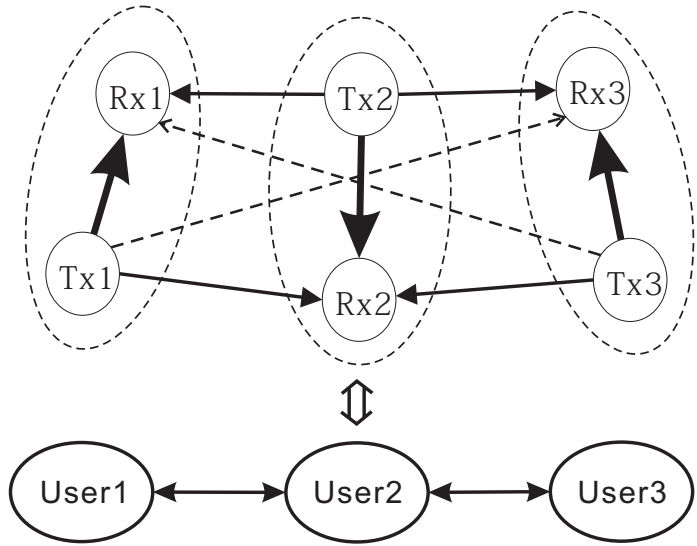} %,height=0.5\linewidth
\caption{3 Tx-Rx Pairs and the corresponding Contention Graph\cite{AlohaGamesSpatialReuse}. In the upper part of the figure, the solid-thick arrow represents the transmission link from a transmitter to its designated receiver; the solid-thin and the dash-thin arrows represent the non-negligible and negligible interference, respectively.} \label{pair}
\vspace{-1em}
\end{figure}

Assume there are $N$ D2D links in the network sharing a dedicated inband overlay channel via CSMA-like random access. These D2D links can transmit in the same frequency band simultaneously if they do not cause any performance degradation to each other. We assume that the CSMA network is hidden-node-free, which can be achieved by properly setting the carrier-sensing power threshold as in \cite{HiddenNodeTVT}\cite{HiddenNodeFreeCS}.
Such a spatial reuse model can be characterized by a ``contention graph" as in \cite{LiewBoE}. For simplicity, only a connected network is considered, and if the network is not connected, then it can be divided into several independent connected sub-networks and dealt with separately. We assume that the contention graph is un-directed and the transmission queue of each D2D link is continuously backlogged, i.e., the transmitter of every D2D link always has a packet to transmit to its designated receiver. An example for three D2D links is shown in Fig. \ref{pair}, where D2D links 1 and 3 can transmit concurrently without collisions but neither of them can transmit together with link 2. In such cases, link 2 is a neighbor of link 1, but link 3 is not. 

\subsection{Ideal CSMA Network Model}

In the CSMA random access method, a link senses the channel before transmitting. Based on such a carrier-sensing relationship, a link will refrain from transmitting if any of its neighbors is transmitting. 
%A link sees the channel as idle if none of its neighbors is transmitting. 
In the ICN model, each link maintains a countdown timer, whose value $t_{cd}$ is modelled as a continuous random variable with an arbitrary distribution\cite{LiewBoE}. The timer value $t_{cd}$ counts down if the channel is sensed as idle, and is frozen if the channel is sensed as busy. When the channel becomes idle again, the countdown of $t_{cd}$ resumes until $t_{cd}=0$, upon which the link transmits a packet. The transmission time $t_{tr}$ is a random variable with an arbitrary distribution. For simplicity, we have adopted uniform distributions for both $t_{cd}$ and  $t_{tr}$ in our simulations.

At any time, a link is either transmitting or idle. Denote the state of link $i$ as $s_i\in \{0, 1\}$, where $s_i=1$ if link $i$ is transmitting and $s_i=0$ otherwise. When $s_i=0$, link $i$ is either actively counting down or frozen, depending on whether a neighboring link $j$ is transmitting or not. We shall denote the system state of a $N$-link ICN by a $N$-tuple binary vector $\vec{s}=[s_1,s_2,\cdots,s_N]$ or simply by a string $s_1s_2\cdots s_N$. Notice that $s_i=s_j=1$ is not allowed if links $i$ and $j$ are neighbors, for the reasons that they can sense each other and the probability of them counting down to zero simultaneously is negligibly small under ICN due to the adopted continuous random variables\cite{LiewBoE}. Therefore, each feasible state corresponds to an independent set\cite{LiewBoE} of the contention graph. 

For the example in Fig. \ref{pair}, the five independent sets are $\O$, $\{1\},\{2\},\{3\},\{1,3\}$. By default, we also include $\O$, which corresponds to $\vec{s}=\vec{0}$, as an independent set. The collection of these feasible system states are denoted by the set 
\begin{equation}\label{SystemStates}
\mathcal{S}=\{[0,0,0],[1,0,0],[0,1,0],[0,0,1],[1,0,1]\}.
\end{equation}%=\{\vec{s}^1,\vec{s}^2,\vec{s}^3,\vec{s}^4,\vec{s}^5\}
If we denote the state $\vec{s}$ with $s_j=0,\forall j\in\mathcal{N}=\{1,2,\cdots,N\}$ as $\vec{e}_0$, and the state $\vec{s}$ with $s_i=1, s_j=0, \forall j\neq i$ as $\vec{e}_i$, then $\mathcal{S}$ can be denoted as 
\begin{equation}\label{SystemStatesE}
\mathcal{S}=\{\vec{e}_0,\vec{e}_1,\vec{e}_2,\vec{e}_3,\vec{e}_1+\vec{e}_3\}.
\end{equation}

\subsection{Stationary Distribution}
\begin{figure}[t]
\centering
\includegraphics[width=0.6\linewidth,  trim=0 0 0 0,clip]{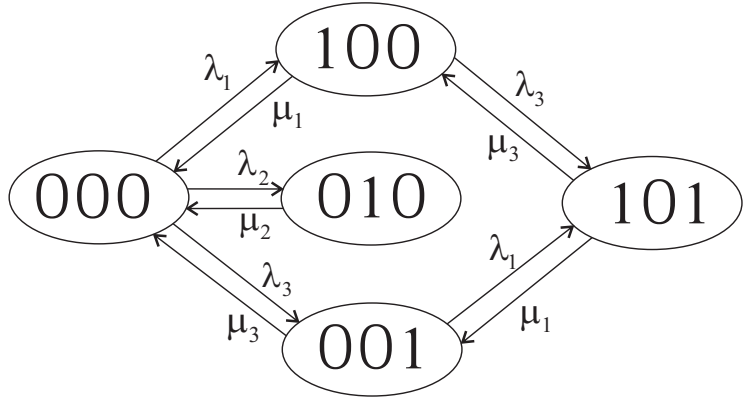} %,height=0.5\linewidth
\caption{State Transition Diagram for Fig. \ref{pair}} \label{Markov3detail}
\vspace{-1em}
\end{figure}

Here we summarize the stationary distribution of the system states based on the results in \cite{LiewBoE}.
If the transmission time and countdown time are exponentially distributed, then the system state $\vec{s}(t)$ is a time-reversible Markov process. 
%Assume that there are $B$ feasible system states, then the set of all feasible system states can be denoted as $\mathcal{S}=\{\vec{s}^1,\vec{s}^2,\cdots,\vec{s}^B\}$.
The state transition diagram of the example in Fig. \ref{pair} is shown in Fig. \ref{Markov3detail}, where there are 5 feasible system states. Each transition from a state in the left to a state in the right represents the beginning of a link transmission,
%, which occurs at a rate $\lambda=1/E[t_{cd}]$;
while the reverse transition represents the ending of the same link transmission.
%, which occurs at a rate $\mu=1/E[t_{tr}]$. 
For example, the transition from 001 to 101 represents the beginning of link 1's transmission while link 3 is transmitting. Similarly, the transition from 101 to 001 represents the ending of link 1's transmission while link 3 continues its transmission. 

The transition rate of a link from idle state to transmission state is $\lambda=1/E[t_{cd}]$, while the transition rate from transmission state to idle state is $\mu=1/E[t_{tr}]$. Hence a higher rate $\lambda$ and a lower rate $\mu$ suggest a higher intensity of the link to access the channel. We define the \textit{access intensity} (AI) \cite{LiewBoE} of a link as the ratio of its mean transmission time to its mean countdown time: $\rho=E[t_{tr}]/E[t_{cd}]=\lambda/\mu$. Note that a higher value of AI suggests a higher intensity to access the channel.
We further define the \textit{transmission aggressiveness} (TA) \cite{JiangLibinCSMA}, which is the natural logarithm of AI $\rho$, i.e., $r=\log_e \rho$. Since natural logarithm is a monotonically increasing function, a higher value of AI corresponds to a higher value of TA, which suggests the link is more aggressive to transmit.

Given a profile of AIs $\vec{\rho}=[\rho_1,\rho_2,\cdots,\rho_N]$, the stationary probability of the state $\vec{s}\in\mathcal{S}$ is shown in \cite{LiewBoE} to be given by
%\begin{equation}\label{DistributionRho}
%P_{\vec{s}^b}=1/Z\cdot\textstyle\prod_{i: s_i^b=1}\rho_i, \forall b=1,\cdots,B,
%\end{equation}%
%where 
%\begin{equation}\label{Zrho}
%Z=\textstyle\sum_{b=1}^B\textstyle\prod_{i: s_i^b=1}\rho_i,
%\end{equation}%
%and by default, $P_{\vec{0}}=1/Z$.
\begin{equation}\label{DistributionRho}
p_{\vec{s}}=1/Z\cdot\textstyle\prod_{i: s_i=1}\rho_i, \forall \vec{s} \in \mathcal{S},
\end{equation}%
where 
\begin{equation}\label{Zrho}
Z=\textstyle\sum_{\vec{s}\in\mathcal{S}}\textstyle\prod_{i: s_i=1}\rho_i.
\end{equation}%
and by default, $p_{\vec{e}_0}=1/Z$. In (\ref{Zrho}), when evaluating $p_{\vec{s}}$, the notation $\prod_{i: s_i=1}\rho_i$ means that for each state $\vec{s}$, only those transmitting links are involved in the multiplication. 
Collectively, we can write the state probability distribution as a vector $\overrightarrow{p}=[p_{\vec{s}_1},p_{\vec{s}_2},\cdots,p_{\vec{s}_{|\mathcal{S}|}}]$, where $|\mathcal{S}|$ is the cardinality of the set $\mathcal{S}$, i.e., the number of feasible states.

Similarly, if we replace AIs by TAs and define a profile of TAs $\vec{r}=[r_1,r_2,\cdots,r_N]=\log_e\vec{\rho}$ for all links, the stationary state probabilities are given by 
\begin{equation}\label{Distribution}
p_{\vec{s}}=1/Z\cdot\exp(\textstyle\sum_{i=1}^N s_i r_i), \forall \vec{s} \in \mathcal{S},
\end{equation}%
where 
\begin{equation}\label{Z}
Z=\textstyle\sum_{\vec{s}\in\mathcal{S}}\exp(\textstyle\sum_{i=1}^N s_i r_i).
\end{equation}%

As an illustration, consider the state transition diagram in Fig. \ref{Markov3detail}. Since the system state is a time-reversible Markov process, the stationary probability distribution should satisfy
\begin{equation}\label{MarkovCompute}
\left\{
    \begin{array}{l l l l l}
      p_{100}=\rho_1\cdot p_{000},\\
      p_{010}=\rho_2\cdot p_{000},\\
      p_{001}=\rho_3\cdot p_{000},\\
      p_{101}=\rho_1\cdot p_{001}=\rho_3\cdot p_{100}=\rho_1\cdot \rho_3\cdot p_{000},\\
      p_{000}+p_{100}+p_{010}+p_{001}+p_{101}=1.
    \end{array} \right.
\end{equation}%
Solving the equations in (\ref{MarkovCompute}) yields
\begin{equation}\label{P000}
p_{000}=1/(1+\rho_1+\rho_2+\rho_3+\rho_1\rho_3)=1/Z,
\end{equation}%
where $Z$ is given in (\ref{Zrho}). Once $Z$ is evaluated, other state probabilities %in (\ref{DistributionRho}) 
can be easily computed. 

Despite the idealized assumption about instantaneous sensing and continuous backoff time, the ICN model does capture the essence of CSMA under spatial reuse. It is shown in \cite{LiewBoE} that the stationary probability distribution in (\ref{DistributionRho}) holds even if both the transmission time and countdown time are not exponentially distributed, given that the ratio of their mean $\rho_i=E[t_{tr,i}]/E[t_{cd,i}]$ for each link $i\in\mathcal{N}$ remains unchanged. On the other hand, in the discrete time model, the stationary probability distribution will deviate from (\ref{DistributionRho}) due to collisions. Fortunately, when RTS/CTS handshaking is used and under the same TA, the stationary distribution will approach that given in (\ref{DistributionRho}) since the collision period will be comparatively small for a sufficiently large holding time \cite{JiangCSMAcollision}.

Finally, it then follows from (\ref{Distribution}) and (\ref{Z}) that the throughput or mean service rate of link $i$ is given by
\begin{equation}\label{throughput}
\theta_i=\textstyle\sum_{\vec{s}\in\mathcal{S}} s_i p_{\vec{s}}=\frac{\textstyle\sum_{\vec{s}\in\mathcal{S}} s_i \exp(\textstyle\sum_{i=1}^N s_i r_i)}{\textstyle\sum_{\vec{s}\in\mathcal{S}}\exp(\textstyle\sum_{i=1}^N s_i r_i)}, \forall i\in\mathcal{N},
\end{equation}%
which is the sum of the stationary state probabilities defined in (\ref{Distribution}) in which link $i$ is actively transmitting (i.e., $s_i=1$). In vector form, if we define the vector $\vec{\theta}=[\theta_1, \theta_2, \cdots, \theta_N]$, then the $N$ equations in (\ref{throughput}) can be collectively written as
\begin{equation}
\vec{\theta}=\textstyle\sum_{\vec{s}\in\mathcal{S}}p_{\vec{s}}\vec{s}.
\end{equation} 
where $\vec{s}$ is the $N$-dimensional vector used to represent a system state.

\section{Feasible Throughput Region in Spatial CSMA Networks}\label{SectionFeasibleThroughput}

In this section, we state and derive the key results on ICN which are important to our proposed game theoretic framework to be presented in Section \ref{Stackelberg}.
 
%Since neighboring links cannot successfully transmit a packet simultaneously, a system state $\vec{s}$ is called feasible if $s_i+s_j\leq 1$, for all neighboring links $i$ and $j$.

\subsection{Feasible and Strictly Feasible Throughput Region}

Each feasible system state $\vec{s}\in \mathcal{S}$ corresponds to a feasible scheduling vector of link transmissions. The feasible throughput region is therefore the convex hull \cite[pp. 24]{Boyd} of $\mathcal{S}$, namely,
\begin{equation}\label{ConvexHull}
\bar{\mathcal{C}}=\{\vec{\theta}|(\vec{\theta}=\sum_{\vec{s}\in\mathcal{S}}p_{\vec{s}}\vec{s})\wedge (p_{\vec{s}}\geq 0, \forall {\vec{s}}) \wedge (\sum_{\vec{s}\in\mathcal{S}} p_{\vec{s}}=1) \}.
\end{equation}%
Eq.(\ref{ConvexHull}) shows that the feasible solutions are given by the convex combinations of the throughputs at these feasible states while fulfilling the probability and probability distribution constraints. The solutions are fully defined by a polytope whose vertices are the feasible system states $\vec{s}\in \mathcal{S}$. 

The interior of $\bar{\mathcal{C}}$ is the \textit{strictly feasible region} $\mathcal{C}$:
\begin{equation}\label{StrictFeasible}
\mathcal{C}=\{\vec{\theta}|(\vec{\theta}=\sum_{\vec{s}\in\mathcal{S}}p_{\vec{s}}\vec{s})\wedge (p_{\vec{s}}> 0, \forall {\vec{s}}) \wedge (\sum_{\vec{s}\in\mathcal{S}} p_{\vec{s}}=1) \}.
\end{equation}%

\begin{figure}[t]
\centering
\includegraphics[width=0.85\linewidth,  trim=0 0 0 0,clip]{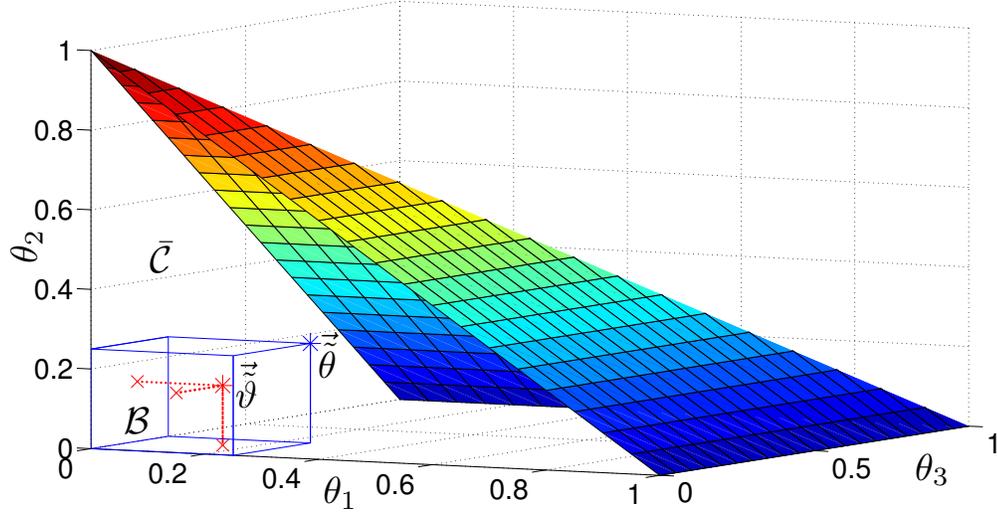} %height=0.5\linewidth
\caption{Feasible Throughput Region for Contention Graph in Fig. \ref{pair}} \label{CSMAfeasibleRegion}
\vspace{-1em}
\end{figure}

Using the contention graph in Fig. \ref{pair} as an example. The set of feasible system states $\mathcal{S}$ has been given in (\ref{SystemStates}). The feasible throughput region is $\bar{\mathcal{C}}$ shown in Fig. \ref{CSMAfeasibleRegion}, which is a polyhedron vertexed by the maximum throughput of these states (the region enclosed by the mesh surface and its intersections with $\theta_1-\theta_2$, $\theta_1-\theta_3$ and $\theta_2-\theta_3$ planes). The strictly feasible region $\mathcal{C}$ refers to the inner region of the polyhedron only. 

%40 20 40 20

\subsection{Transmission Aggressiveness}\label{SectionLemma}
CSMA is a distributed and randomized way to schedule the transmissions among the feasible system states.
It is shown in \cite{JiangLibinCSMA} using the ICN model that any throughput in the strictly feasible region can be achieved through a properly chosen TA $\vec{r}$, which is stated in the following lemma.

\newtheorem{lemma}{Lemma}
\begin{lemma}[Lemma 8 in \cite{ICNuniqueProof}]\label{CSMAlemma}
In the ICN model, for any desired throughput for all the $N$ links $\vec{\tilde{\theta}}=[\tilde{\theta}_1, \tilde{\theta}_2, \cdots, \tilde{\theta}_N] \in \mathcal{C}$ (strictly feasible region), there exists a unique finite-valued $\vec{r}=[r_1, r_2, \cdots, r_N]\in \mathcal{R}^N$ such that $\theta_i(\vec{r})= \tilde{\theta}_i, \forall i\in \mathcal{N}$.
\end{lemma}%$\log(\rho_i)\geq 0, \forall i\in \mathcal{N}$

A detailed proof can be found in \cite{JiangLibinCSMA} and \cite{ICNuniqueProof}. Here we only present a sketch of the proof.

\textit{Proof}: Given a $\vec{\tilde{\theta}}\in \mathcal{C}$, we use the maximum log-likelihood method to estimate the parameters $\vec{r}^*$ which result in $\vec{\theta}(\vec{r}^*)=\vec{\tilde{\theta}}$, or equivalently, result in the desired state probability distribution $\overrightarrow{p}^{\tilde{\theta}}$ such that
$\vec{\theta}(\vec{r}^*)=\textstyle\sum_{\vec{s}\in\mathcal{S}}p_{\vec{s}}^{\tilde{\theta}}\vec{s}$.
The log-likelihood function \cite{LogLikelihood} is defined as:
\begin{equation}\label{LogLikelihoodOrigin}
F(\vec{r};\vec{\tilde{\theta}})=\sum_{\vec{s}\in\mathcal{S}}p_{\vec{s}}^{\tilde{\theta}}\log_e(p_{\vec{s}}).
\end{equation}%
By applying $\vec{\tilde{\theta}}=\textstyle\sum_{\vec{s}\in\mathcal{S}}p_{\vec{s}}^{\tilde{\theta}}\vec{s}$ and substituting the expression for $p_{\vec{s}}$ given in (\ref{Distribution}), and after some manipulations, we have
\begin{equation}\label{LogLikelihood}
F(\vec{r};\vec{\tilde{\theta}})=\textstyle\sum_{i=1}^N \tilde{\theta}_i r_i-\log_e [\textstyle\sum_{\vec{s}\in\mathcal{S}}\exp(\textstyle\sum_{i=1}^N s_i r_i)].
\end{equation}%
Since $\textstyle\sum_{i=1}^N \tilde{\theta}_i r_i$ is affine in $\vec{r}$ and $\log_e [\textstyle\sum_{\vec{s}\in\mathcal{S}}\exp(\textstyle\sum_{i=1}^N s_i r_i)]$ is a log-sum-exp function and thus is convex in $\vec{r}$,
the function $F(\vec{r};\vec{\tilde{\theta}})$ is concave in $\vec{r}$ \cite[pp. 72]{Boyd}.
Therefore, the max-log-likelihood problem below is a convex optimization problem with $\vec{r}$ as the variables to be solved and $\vec{\tilde{\theta}}$ as the parameters:
\begin{equation}\label{MaxLikelihood1}
\max\limits_{\vec{r}} ~~F(\vec{r};\vec{\tilde{\theta}})\quad \textrm{(Maximize log-likelihood)}.
\end{equation}

It is then shown in \cite{JiangLibinCSMA} that the max-log-likelihood problem in (\ref{MaxLikelihood1}) is the dual problem of the max-entropy problem in (\ref{MaxEntropy1}), where $-\textstyle\sum_{\vec{s}\in\mathcal{S}} p_{\vec{s}}\log_e p_{\vec{s}}$ is the entropy of the distribution vector $\overrightarrow{p}$, whose element $p_{\vec{s}}$ is the state probability for the state $\vec{s}, \forall \vec{s}\in\mathcal{S}$. The max-entropy problem is also a convex optimization problem, with $\overrightarrow{p}$ as the variables and $\vec{\tilde{\theta}}$ as the parameters.

\begin{equation}\label{MaxEntropy1}
\begin{array}{l l}
\max\limits_{\overrightarrow{p}} ~~ -\textstyle\sum_{\vec{s}\in\mathcal{S}} p_{\vec{s}}\log_e p_{\vec{s}} \quad \textrm{(Maximize entropy)}\\
\mbox{s.t.} ~~\left\{
    \begin{array}{l l l}
      \textstyle\sum_{\vec{s}\in\mathcal{S}} s_i p_{\vec{s}}= \tilde{\theta}_i, \forall i\in \mathcal{N},\\
      p_{\vec{s}}\geq 0, \forall \vec{s}\in\mathcal{S},\\
      \textstyle\sum_{\vec{s}\in\mathcal{S}} p_{\vec{s}}=1.\\
    \end{array} \right.
\end{array}
\end{equation}%

We are now ready to prove Lemma \ref{CSMAlemma}.
We need to verify that the Slater's condition \cite[pp. 226]{Boyd} is satisfied, so that the optimal solutions to the two convex optimization problems (\ref{MaxLikelihood1}) (\ref{MaxEntropy1}) exist with zero duality gap, given that $\vec{\tilde{\theta}}\in \mathcal{C}$ (strictly feasible region).

Since all the constraints in (\ref{MaxEntropy1}) are linear equalities and inequalities, we only need to verify that there exists a feasible $\overrightarrow{p}$ in the relative interior \cite[pp. 23]{Boyd} of the domain $\mathcal{D}$ of the objective function $-\textstyle\sum_{\vec{s}\in\mathcal{S}} p_{\vec{s}}\log_e p_{\vec{s}}$, which is $\mathcal{D}=\{\overrightarrow{p} | p_{\vec{s}}\geq 0, \forall \vec{s}\in\mathcal{S}\}$.
The relative interior of $\mathcal{D}$ is $\textbf{relint} \mathcal{D}=\{\overrightarrow{p} | p_{\vec{s}}> 0, \forall \vec{s}\in\mathcal{S}\}$.
Since $\vec{\tilde{\theta}}\in \mathcal{C}$, from (\ref{StrictFeasible}) we can write $\vec{\tilde{\theta}}=\textstyle\sum_{\vec{s}\in\mathcal{S}}p_{\vec{s}}^{\tilde{\theta}}\vec{s}$
where $p_{\vec{s}}^{\tilde{\theta}}>0,\forall \vec{s}\in\mathcal{S}$ and $\sum_{\vec{s}\in\mathcal{S}}p_{\vec{s}}^{\tilde{\theta}}=1$.
By letting $\overrightarrow{p}=\overrightarrow{p}^{\tilde{\theta}}\in \textbf{relint} \mathcal{D}$, we find a feasible $\overrightarrow{p}$ which satisfies all the constraints in (\ref{MaxEntropy1}). Therefore, the Slater's condition is satisfied.

As a result, the optimal solutions to the two convex optimization problems (\ref{MaxLikelihood1}) (\ref{MaxEntropy1}) exist with zero duality gap. Moreover, the dual optimal value is attainable, i.e., there exists a finite $\vec{r}^*$ such that $F(\vec{r}^*;\vec{\tilde{\theta}})=\max_{\vec{r}} F(\vec{r};\vec{\tilde{\theta}})$.
Therefore, the first order condition \cite[pp. 457]{Boyd} of the unconstrainted differentiable convex optimization problem in (\ref{MaxLikelihood1}) is satisfied at $\vec{r}^*$, i.e.,
\begin{equation}\label{FirstOrder}
\nabla F(\vec{r};\vec{\tilde{\theta}})\mid_{\vec{r}=\vec{r}^*}=\vec{0},
\end{equation}%
which yields
\begin{multline}\label{FirtOrderPartial}
\frac{\partial F(\vec{r};\vec{\tilde{\theta}})}{\partial r_i}\mid_{\vec{r}=\vec{r}^*}=\tilde{\theta}_i-\frac{\textstyle\sum_{\vec{s}\in\mathcal{S}} s_i \exp(\textstyle\sum_{i=1}^N s_i r_i^*)}{\textstyle\sum_{\vec{s}\in\mathcal{S}}\exp(\textstyle\sum_{i=1}^N s_i r_i^*)}\\
=\tilde{\theta}_i-\textstyle\sum_{\vec{s}\in\mathcal{S}} s_i p_{\vec{s}}=\tilde{\theta}_i-\theta_i^*=0, \forall i\in \mathcal{N}.
\end{multline}%

Therefore, for any $\vec{\tilde{\theta}}\in \mathcal{C}$ (strictly feasible region), 
the log-likelihood function
$F(\vec{r};\vec{\tilde{\theta}})$ attains its maximum value at a finite-valued $\vec{r}=\vec{r}^*\in \mathcal{R}^N$. At the optimal solution $\vec{r}^*$, the first-order optimality condition (\ref{FirstOrder}) is satisfied, which corresponds to $\theta_i^*(\vec{r}^*)= \tilde{\theta}_i, \forall i\in \mathcal{N}$.
It is further shown in \cite{ICNuniqueProof} that $F(\vec{r};\vec{\tilde{\theta}})$ is strictly concave in $\vec{r}$. Therefore, the optimal solution $\vec{r}^*$ is unique.
$\blacksquare$

Lemma \ref{CSMAlemma} suggests that, if $\vec{\tilde{\theta}}\in \mathcal{C}$, then a unique solution $\vec{r}^*$ exists such that $\theta_i^*(\vec{r}^*)= \tilde{\theta}_i, \forall i\in \mathcal{N}$. The above proof also suggests that we can solve for $\vec{r}^*$ by maximizing the concave function $F(\vec{r};\vec{\tilde{\theta}})$. This is useful for the design of our game iteration algorithm presented in Section \ref{CSMAgame}.

\subsection{Feasible Throughput Region Under ICN}
Previously we have defined the feasible throughput region for any given set of feasible system states. The shape of the polytopes derived from the ICN model owns a property which will be discussed here.

We first introduce a binary relation ``$ \preceq $" between two real-valued vectors $\vec{\tilde{\vartheta}}$ and $
\vec{\tilde{\theta}}$, which is defined as component-wise less than or equal to, i.e., 
\begin{equation}
\vec{\tilde{\vartheta}}\preceq \vec{\tilde{\theta}}\Leftrightarrow \tilde{\vartheta}_i\leq \tilde{\theta}_i,\forall i\in \mathcal{N}. 
\end{equation}%
We now establish the following theorem which will be useful when presenting our proposed games.  

\newtheorem{theorem}{Theorem}
\begin{theorem}\label{PartialOrderTheorem}
In the ICN model, given that $\vec{\tilde{\theta}}\in \bar{\mathcal{C}}$ ($\tilde{\theta}$ is in the feasible region), then any desired throughput $\vec{\tilde{\vartheta}}$, where $\vec{0}\preceq\vec{\tilde{\vartheta}}\preceq \vec{\tilde{\theta}}$, is also in $\bar{\mathcal{C}}$.
\end{theorem}%$\log(\rho_i)\geq 0, \forall i\in \mathcal{N}$

\begin{figure}[t]
        \centering
        \begin{subfigure}[b]{0.5\linewidth}
                \includegraphics[width=1\linewidth,  trim=0 0 0 0,clip]{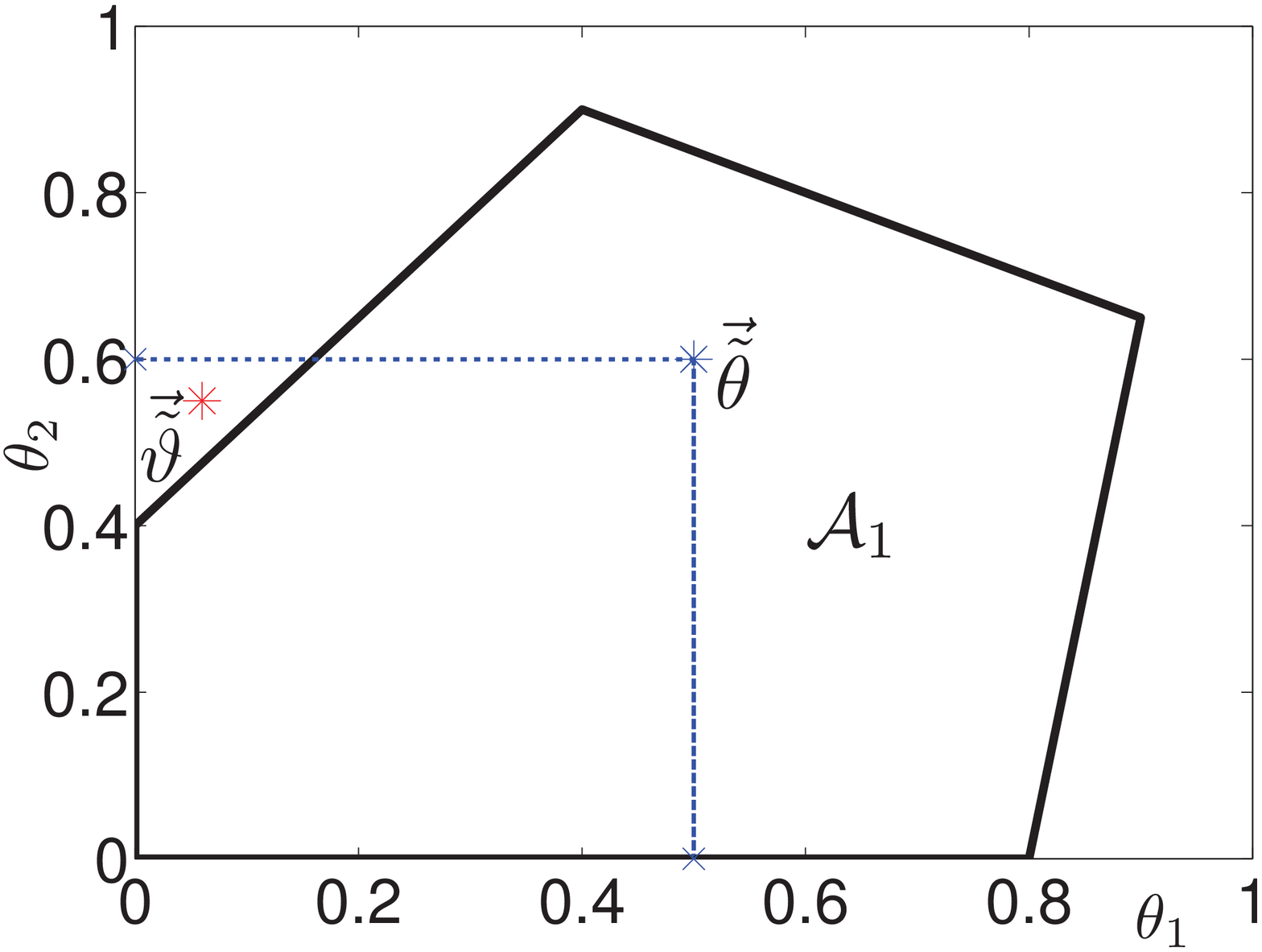}
                \caption{Convex Polytope $\mathcal{A}_1$ without Property in Theorem \ref{PartialOrderTheorem}}
                \label{ConvexSetNotPartialOrder}
        \end{subfigure}%
        ~ %add desired spacing between images, e. g. ~, \quad, \qquad, \hfill etc.
        \begin{subfigure}[b]{0.5\linewidth}
                \includegraphics[width=1\linewidth,  trim=0 0 0 0,clip]{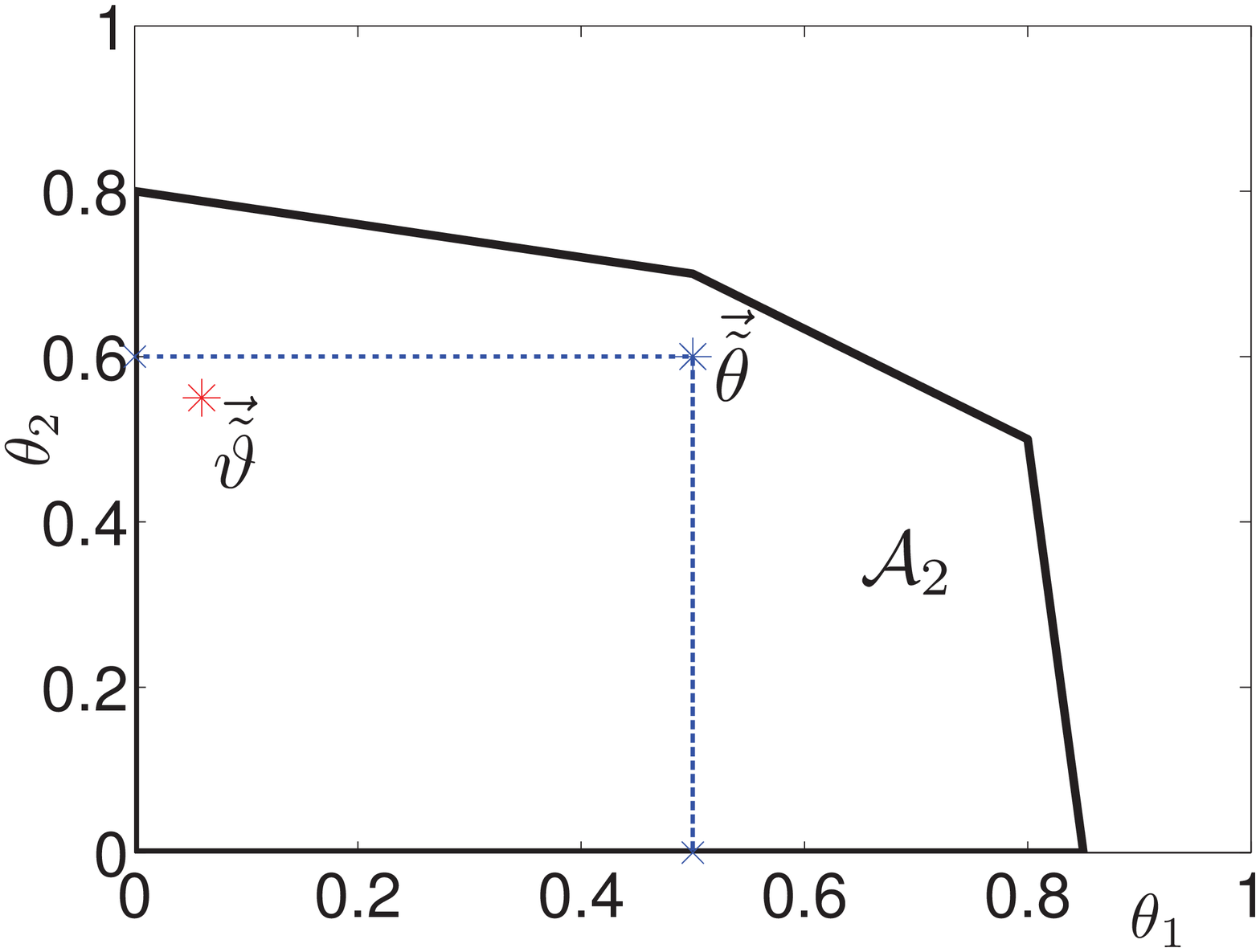}
                \caption{Convex Polytope $\mathcal{A}_2$ with Property in Theorem \ref{PartialOrderTheorem}}
                \label{ConvexSetPartialOrder}
        \end{subfigure}% 
        \caption{Examples of 2-Dimensional Convex ``Polytopes"}\label{Polytope}
        %\vspace{-1.5em}
\end{figure}

\textit{Proof}: 
A first glance at Fig. \ref{CSMAfeasibleRegion} may lead to the thought that the theorem is trivial, but this is not true. Fig. \ref{ConvexSetNotPartialOrder} shows a convex set $\mathcal{A}_1$ %(the black solid lines and the region within) 
in the two-dimensional space. For a $\vec{\tilde{\theta}}\in \mathcal{A}_1$ as shown in Fig. \ref{ConvexSetNotPartialOrder}, it is easy to find a point $\vec{\tilde{\vartheta}}$ such that $\vec{\tilde{\vartheta}}\preceq \vec{\tilde{\theta}}$ and yet $\vec{\tilde{\vartheta}}$ is not within the convex region $\mathcal{A}_1$. On the other hand, it is not difficult to figure out that the convex set $\mathcal{A}_2$ in Fig. \ref{ConvexSetPartialOrder} owns the property stated in Theorem \ref{PartialOrderTheorem}.

In the ICN model, for a target throughput vector $\vec{\tilde{\theta}}$ where $\vec{\tilde{\theta}}\in \bar{\mathcal{C}}$, there exists a probability distribution $\overrightarrow{p}^{\tilde{\theta}} = \{p_{\vec{s}}^{\tilde{\theta}}, \forall \vec{s} \in \mathcal{S}\}$ where $\vec{\tilde{\theta}}=\sum_{\vec{s}\in\mathcal{S}}p_{\vec{s}}^{\tilde{\theta}}\vec{s}$ according to (\ref{ConvexHull}).
%
%$\vec{p}^{\tilde{\theta}}=\{p_{\vec{s}}^{\tilde{\theta}}|p_{\vec{s}}^{\tilde{\theta}}\geq 0, \forall \vec{s}\in\mathcal{S}; \sum_{\vec{s}\in\mathcal{S}} p_{\vec{s}}^{\tilde{\theta}}=1\}$ such that $\vec{\tilde{\theta}}$ can be expressed as $\vec{\tilde{\theta}}=\sum_{\vec{s}\in\mathcal{S}}p_{\vec{s}}^{\tilde{\theta}}\vec{s}$, according to (\ref{ConvexHull}).
%
To prove that $\vec{\tilde{\vartheta}} \preceq \vec{\tilde{\theta}} \in \bar{\mathcal{C}}$, we need to similarly show that there exists another probability distribution $\overrightarrow{p}^{\tilde{\vartheta}} = \{ p_{\vec{s}}^{\tilde{\vartheta}}, \forall \vec{s}\in \mathcal{S} \}$ that fulfills (\ref{ConvexHull}).
However, it is difficult to obtain the distribution $\overrightarrow{p}^{\tilde{\vartheta}}$ directly from $\overrightarrow{p}^{\tilde{\theta}}$ since it depends on the underlying link topology.

Our approach is to define an orthotope $\mathcal{B}$
%=\{\vec{\tilde{\vartheta}}|\vec{0}\preceq\vec{\tilde{\vartheta}}\preceq\vec{\tilde{\theta}}\}$ %beneath $\vec{\tilde{\theta}}$, 
whose ``vertices" are obtained by projecting $\vec{\tilde{\theta}}$ on all the coordinate planes. An example for the 3-dimensional illustration is shown in Fig. \ref{CSMAfeasibleRegion}. The problem is now becoming equivalent to showing that all the ``vertices" of $\mathcal{B}$ are in $\bar{\mathcal{C}}$. Finally, because $\vec{\tilde{\vartheta}}\preceq \vec{\tilde{\theta}}$, $\vec{\tilde{\vartheta}}$ is within the cuboid and hence within $\bar{\mathcal{C}}$.

We first perform a projection parallel to the $i$-th axis. Consider a throughput vector $\vec{\tilde{\psi}}$, with the setting of $\tilde{\psi}_i=0,\tilde{\psi}_j=\tilde{\theta}_j, \forall j\neq i$, i.e., the $i$-th link has zero throughput. %Clearly $\vec{\tilde{\psi}}$ is the projection of $\vec{\tilde{\theta}}$ onto the coordinate plane where the $i-$th dimension takes a value of 0. 
It is intuitive that $\vec{\tilde{\psi}}$ is one of the ``vertices" of $\mathcal{B}$. In order to show that $\vec{\tilde{\psi}}$ is in $\bar{\mathcal{C}}$, we need to show that we are able to obtain its state probability distribution $\{\overrightarrow{p}^{\tilde{\psi}}\}$ from $\{\overrightarrow{p}^{\tilde{\theta}}\}$, and $\vec{\tilde{\psi}}$ can be expressed in the form as in (\ref{ConvexHull}). This can be done in the following way.

For $\vec{\tilde{\theta}}\in \bar{\mathcal{C}}$, its state distribution $\overrightarrow{p}^{\tilde{\theta}}$ satisfies $\vec{\tilde{\theta}}=\sum_{\vec{s}\in\mathcal{S}}p_{\vec{s}}^{\tilde{\theta}}\vec{s}$, $p_{\vec{s}}^{\tilde{\theta}}\geq 0,\forall \vec{s}\in\mathcal{S}$ and $\sum_{\vec{s}\in\mathcal{S}} p_{\vec{s}}^{\tilde{\theta}}=1$. 
%We call $\vec{u}^{\theta}=\{u_{\vec{s}}^{\theta}|\vec{s}\in\mathcal{S}\}$ the state distribution which expressed $\vec{\tilde{\theta}}$.
We next describe how to construct the state distribution $\overrightarrow{p}^{\tilde{\psi}}$ for $\vec{\tilde{\psi}}$.
For those states in $\mathcal{S}$ with $s_i=1$, choose $p_{\vec{s}}^{\tilde{\psi}}=0$ and $p_{\vec{s}-\vec{e}_i}^{\tilde{\psi}}=p_{\vec{s}-\vec{e}_i}^{\tilde{\theta}}+p_{\vec{s}}^{\tilde{\theta}}$. 
For the remaining states, choose $p_{\vec{s}}^{\tilde{\psi}}=p_{\vec{s}}^{\tilde{\theta}}$. In other words, those states $\vec{s}$ with $s_i=1$ should now have state probability $p_{\vec{s}}^{\tilde{\psi}}=0$. The ``removed" state probability $p_{\vec{s}}^{\tilde{\psi}}$ should now be attributed to the state $\vec{s}-\vec{e}_i$. It is not difficult to verify that, by doing so, the total probability remains one and the throughputs of all unaffected links remain the same as before. This state probability distribution $p_{\vec{s}}^{\tilde{\psi}}$ clearly satisfies (\ref{ConvexHull}), hence we conclude that the vertex $\vec{\tilde{\psi}}$ is within $\bar{\mathcal{C}}$ and so are other vertexes of $\mathcal{B}$. 

For the example shown in Fig. \ref{Markov3detail},
assume that we have a throughput $\vec{\tilde{\theta}}=[\tilde{\theta}_1,\tilde{\theta}_2,\tilde{\theta}_3]\in \bar{\mathcal{C}}$. We now show that $\vec{\tilde{\psi}}=[\tilde{\psi}_1,\tilde{\psi}_2,\tilde{\psi}_3]=[0,\tilde{\theta}_2,\tilde{\theta}_3]$ is also in $\bar{\mathcal{C}}$.
Note that the throughput $\vec{\tilde{\psi}}$ is equivalent to the case in which link 1 powers off and stops transmitting. In such a case, there are only three feasible system states left: 000, 010, 001. In other words, the states 100 and 101 disappear and are merged into the states 000 and 001 respectively, since link 1 is no longer transmitting. State 010 remains unchanged.
Merging the state probability $p_{101}$ with $p_{001}$ will ensure the throughput for link 3 remains the same, since $\tilde{\theta}_3=p_{001}+p_{101}$. Merging the state probability $p_{100}$ with $p_{000}$ will not affect the throughput of any remaining links. 
Since the total probability still sum up to be one, link 1 will not be transmitting and both link 2 and link 3 transmit as before. 
Therefore, the throughput $\vec{\tilde{\psi}}$ resulting from the above state merging operations is still in $\bar{\mathcal{C}}$.

Other vertices of $\mathcal{B}$ can also be similarly shown to be in $\bar{\mathcal{C}}$.  Since $\bar{\mathcal{C}}$ is a convex set and the convex combination of these ``vertices" are all in $\bar{\mathcal{C}}$, we have $\mathcal{B}\subset\bar{\mathcal{C}}$. Since $\vec{\tilde{\vartheta}}\preceq \vec{\tilde{\theta}}$ is enclosed in the hyperrectangle ($N$-orthotope), $\vec{\tilde{\vartheta}}$ should also be in $\bar{\mathcal{C}}$.
$\blacksquare$

%Similar proof of results can be obtained for $\mathcal{C}$ as well.
\textit{Remark}: Theorem \ref{PartialOrderTheorem} is not generally true for any convex set. It is true since the values of $s_i$ are chosen from 0 and 1 only; and the subset of feasible states induced by a maximal independent set is a complete partially ordered set\cite{AlohaGamesSpatialReuse} based on how ICN is modelled. For the example in Fig. \ref{Markov3detail}, the maximal independent set $\{1,3\}$ induces the subset of feasible states $\mathcal{Q}=\{[0,0,0],[1,0,0],[0,0,1],[1,0,1]\}$, which is a complete partially ordered set, with the least element $[0,0,0]$ and the largest element $[1,0,1]$ under the partial order ``$\preceq$". Hence the use of the theorem needs to be carefully dealt with.

% The physical meaning of Theorem \ref{PartialOrderTheorem} is quite straightforward...

Theorem \ref{PartialOrderTheorem} will be used in Section \ref{QuasiConvex} to show that the pricing problem is a valid quasi-convex optimization problem.

\subsection{D2D Network Model}\label{D2Dmodel}
This subsection discusses how to efficiently model the resulting D2D network if CSMA is adopted by all D2D links. If the objective of the network is to maximize the sum-rate of all transmitting links, and the BS gives no control on the admission and transmission of links, then \cite{JiangLibinCSMA} has successfully solved this problem. The solution is computed in a completely distributed manner. However, as pointed out earlier, such a fully cooperative model is too idealistic and there is no consideration on the utility heterogeneity and selfish behaviors of the links. It may be better to build a pricing framework so that each link tries to maximize its payoff function when competing for resources, rather than someone tries to take advantage when the network is in operation and drive the network to unstable states. 
Furthermore, maximizing the sum-rate may not distribute the resources according to demand because links with low demand may be assigned to transmit at higher rates due to its spatial location. 

In this paper, we feel that the BS can take a more proactive role to assist in D2D transmission. In fact, the problem can be formulated separately in terms of the objectives of the D2D links and the BS. The objective of the BS is to maximize the sum-rate while satisfying the physical layer constraints:
\begin{equation}\label{BSObjective}
\begin{array}{l l}
\max ~~ \textstyle\sum_{i=1}^N \tilde{\theta}_i\\
\mbox{s.t.} ~~\vec{\tilde{\theta}}\in\mathcal{C},
\end{array}
\end{equation}%
where $\tilde{\theta}_i$ is the target rate the network has to support link $i$, and the solution must fulfill the CSMA channel access constraint, i.e., the final rates to support all D2D links must be in the strictly feasible throughput region $\mathcal{C}$ defined in (\ref{StrictFeasible}).

Each link is a player of a non-cooperative game. Each player tries to maximize its payoff $v_i(\theta_i, \theta_{-i})$ while satisfying the physical layer constraints. 
\begin{equation}\label{D2DObjective}
\begin{array}{l l}
	\max ~~ v_i(\theta_i, \theta_{-i}), \forall i\\
	\mbox{s.t.}  ~~\vec{\theta}\in\mathcal{C}.
\end{array}
\end{equation}%

In (\ref{BSObjective}), $\tilde\theta_i$ is used to represent the rate demand from the utility point of view and should be differentiated from $\theta_i$ in (\ref{D2DObjective}) or ($\ref{throughput}$) used in the ICN model as a result of competing for channel access. At the equilibrium state, these two quantities have to be the same and the pricing mechanism aims to achieve this objective. 

There are two challenges in the formulation. The above two optimization problems both involve the constraint defined by the strictly feasible throughput region $\mathcal{C}$. From Lemma \ref{CSMAlemma} we know that, for any desired throughput $\vec{\tilde{\theta}}$ in the strictly feasible region $\mathcal{C}$, there exists an operating point $\vec{r}$ such that $\vec{\theta}(\vec{r})=\vec{\tilde{\theta}}$. However, in order to obtain $\mathcal{C}$ as in (\ref{StrictFeasible}), we need to know all the feasible system states, which correspond to all the independent sets \cite{LiewBoE} in the contention graph. As is shown in \cite{LiewBoE}, to compute all the independent sets (include the maximal independent sets) is a NP-hard problem. Hence it is practically difficult to obtain $\mathcal{C}$. 
%, especially the underlaying D2D topology is time-varying. 
The second challenge is how to align the solution of (\ref{BSObjective}) with the involvement of (\ref{D2DObjective}). 

Our approach is to develop a simple mechanism which does not require a-prior knowledge of $\mathcal{C}$ and yet the radio resource can be allocated to the heterogeneous D2D links efficiently while satisfying the objectives of both the BS and D2D links. A pricing mechanism is introduced to achieve this purpose. The payoff function of each link is made to be dependent on the resource price. The BS broadcasts the resource price and uses it to control the transmission behavior of each link.  
%, it is unable to decide whether a certain target rate $\vec{\tilde{\theta}}$ can be satisfied, i.e., whether or not $\vec{\tilde{\theta}}\in \mathcal{C}$. Therefore, to design the pricing strategies for the leader so that its objective can be iteratively achieved. 
%The objective of the leader is to choose a price $M$ so that the total throughput of the D2D links is maximized under the constraint that the target rate $\vec{\tilde{\theta}}$ of the heterogeneous users can all be satisfied. From Proposition \ref{csmaNEexistence}, such a constraint is equivalent to $\vec{\tilde{\theta}}\in\mathcal{C}$.
%
Mathematically, the BS solves the following optimization problem:
\begin{equation}\label{LeaderProblemC}
\max\limits_{M\geq 0} ~~g(M):=\textstyle\sum_{i=1}^N \tilde{\theta}_i(M)\\
\end{equation}%
where $\tilde{\theta}_i(M)$ is the target rate of D2D link $i$ under the service price $M$, which will be presented in (\ref{TargetRate}).
The D2D links are the followers in the overall Stackelberg game, each of which chooses its transmission strategy so that its individual payoff is maximized under the service price chosen by the BS, i.e., the Stackelberg leader. 
%In the subgame model in Section \ref{CSMAgame}

In the next section, we describe how our proposed Stackelberg game model can achieve the above purposes.

\section{Stackelberg Games for Non-Cooperative D2D Links}\label{Stackelberg}

Stackelberg games\cite{Fudenberg} are a class of non-cooperative games in which a leader, who makes the first move in the game, anticipates the actions of the followers based on a model of how the followers would respond to its actions. We propose a Stackelberg game, in which the BS in the cellular network acts as a Stackelberg leader to regulate the transmission behaviors of all the D2D links by broadcasting a proper service price $M$. The D2D links are the followers, each of which responds to the price $M$ by choosing its transmission strategy in an attempt to maximize its individual payoff.% so that their individual payoff is maximized. %The D2D links then compete for channel usage by choosing their transmission parameters in order to achieve their desired throughputs.

In Section \ref{UtilityFunction}, we first define the utility functions for the D2D links, each of which characterizes the individual service requirements and willingness to pay. In Section \ref{CSMAgame}, we study the non-cooperative behaviors of the D2D links under a given network price $M$, which defines the follower-subgame in the Stackelberg game.
%The objective of the BS, i.e., the Stackelberg leader, is described in Section \ref{BSobjective}, which is to maximize the sum rate under the constraint that the target rates of all the heterogeneous D2D links can all be satisfied. 
The Stackelberg game is analyzed in Section \ref{QuasiConvex}. Based on the analysis, the pricing strategies of the Stackelberg leader are proposed in Section \ref{PricingStrategies}. A brief complexity analysis is given in Section \ref{SectionComplexity}.
%Then in Part D we extend the game from the ICN model to practical CSMA networks with discrete contention window sizes, and discuss the necessary modifications in the algorithm.
%Finally we comment on the information requirements in the proposed game in Part E.

\subsection{D2D Link Utility Function}\label{UtilityFunction}
We modify the traffic model used in \cite{alohaprice} to our system. Suppose D2D link $i$ has a target rate $\tilde\theta_i$ in the range of $[\gamma_i, \pi_i]$, where $\gamma_i \leq {\tilde\theta}_i \leq \pi_i$. 
If $\tilde\theta_i < \gamma_i$, link $i$ achieves zero utility, and each link has no intention to go beyond $\tilde\theta_i > \pi_i$. The exact target rate value $\tilde\theta_i$ is controlled by the service price $M$ through the following relationship
\begin{equation}\label{TargetRate}
\begin{array}{l l}
	\tilde\theta_i(M)=\left\{
    \begin{array}{l l}
      0, &\textrm{ $M>m_i$,}\\
      \min\{\gamma_i-b_i(M-m_i),\pi_i\}, & \textrm{ $0\leq M\leq m_i$},\\
    \end{array} \right.
\end{array}
\end{equation}
where $\gamma_i$, $\pi_i$
%is another factor enforced by the BS to control users' adapting rate with the price $M$, 
and $m_i$ together decide how link $i$ is willing to pay for the transmission. For simplicity, we have adopted a monotonically decreasing linear function for $\tilde\theta_i(M)$ in the range $\gamma_i \leq {\tilde\theta}_i \leq \pi_i$, where $b_i$ is a positive coefficient and $-b_i$ is the slope. 

Eq. (\ref{TargetRate}) is interpreted as follows. The parameter $m_i$ is the highest price that link $i$ is willing to pay for its transmission. When $M=m_i$, link $i$ will only desire a minimum throughput of $\gamma_i$. When the price is too high (i.e., $M>m_i$), link $i$ chooses not to transmit, and thus its target rate drops to zero, i.e., $\tilde{\theta}_i(M)=0$. Over the range $0\leq M\leq m_i$, link $i$ is willing to pay for its transmission, and the lower the price $M$, the higher throughput it desires, unless it has already reached its maximum desired throughput $\pi_i$. 
In this range, we have used a linear function to simplify the above monotonic relationship. Other function forms such as hyperbolic, parabolic, cubic $\cdots$ can also be used, as long as the monotonic relationship is preserved.
As a result, the relationship in (\ref{TargetRate}) is a piecewise linear function. In the special case where the minimum desired throughput is $\gamma_i=0$ and the maximum desired throughput $\pi_i\geq b_i m_i$, the piecewise linear relationship in (\ref{TargetRate}) simply reduces to a smooth linear relationship.
A smooth monotonic curve can be similarly obtained when other function forms are adopted. Note that our algorithm works if only the monotonic function property holds.

%Also note that the relationship in (\ref{TargetRate}) include the case 

%the target rate $\tilde{\theta}_i(M)$ is increasing when the price $M$ is decreasing, until reaching its maximum desired throughput, i.e., $\tilde{\theta}_i(M)=\pi_i$. In other words, link $i$ will not be greedy for throughput higher than $\pi_i$, and on the other hand, each link chooses not to transmit if the price is too high.  

%The relationship in (\ref{TargetRate}) can be illustrated using the three links' example in Fig. \ref{TargetRate3}. It can be seen that the target rate curve is a piecewise linear function of the service price $M$. For example, link 3 has a minimum desired throughput $\gamma_3=0.1$, When the price $M$ is higher than 

\begin{figure}[t]
        \centering
        \begin{subfigure}[b]{0.5\linewidth}
                \includegraphics[width=1\linewidth,  trim=0 0 0 0,clip]{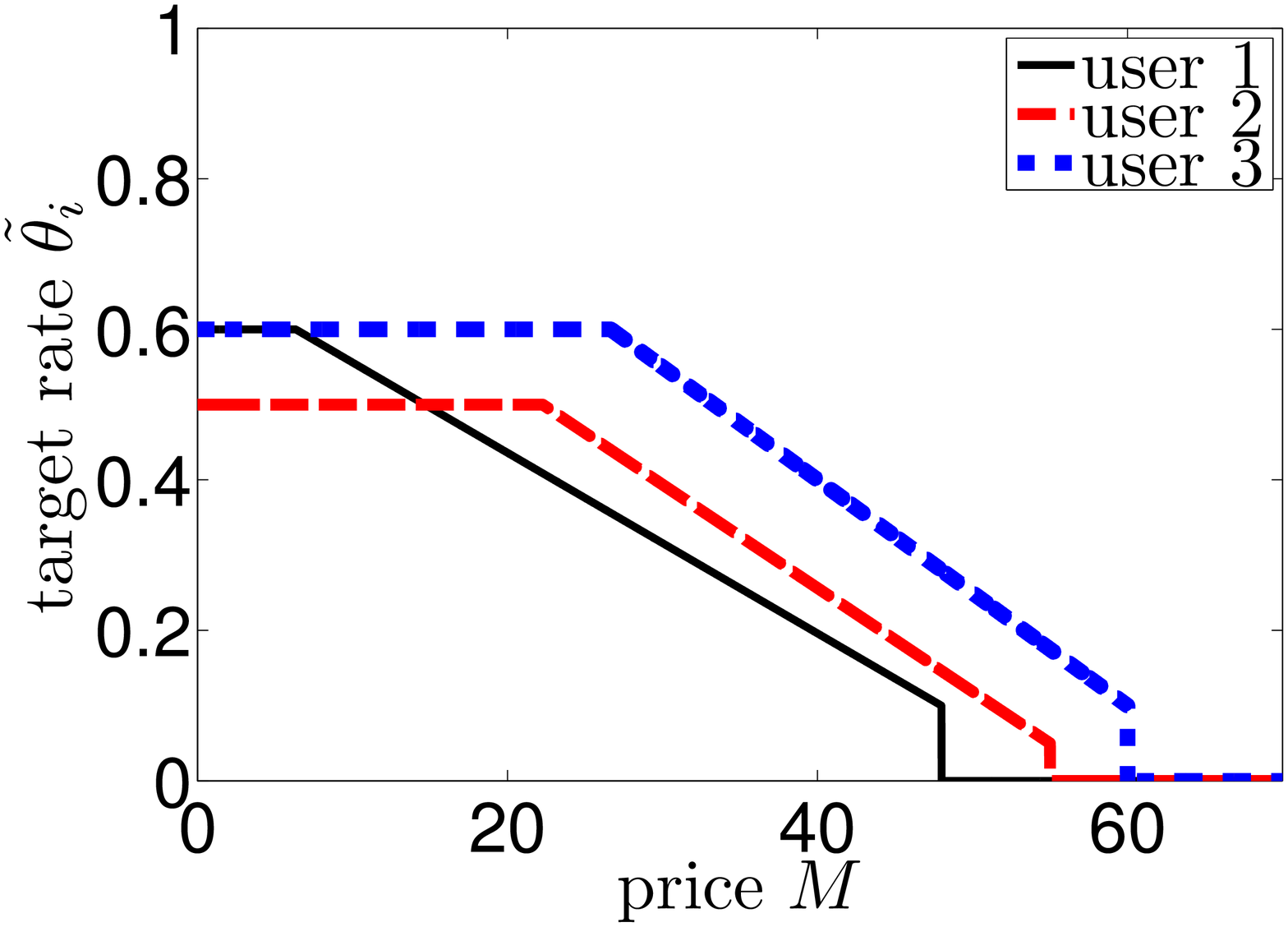}
                \caption{Target Rates of the CSMA Users under the Price $M$}
                \label{TargetRate3}
        \end{subfigure}%
        ~ %add desired spacing between images, e. g. ~, \quad, \qquad, \hfill etc.
        \begin{subfigure}[b]{0.5\linewidth}
                \includegraphics[width=1\linewidth,  trim=0 0 0 0,clip]{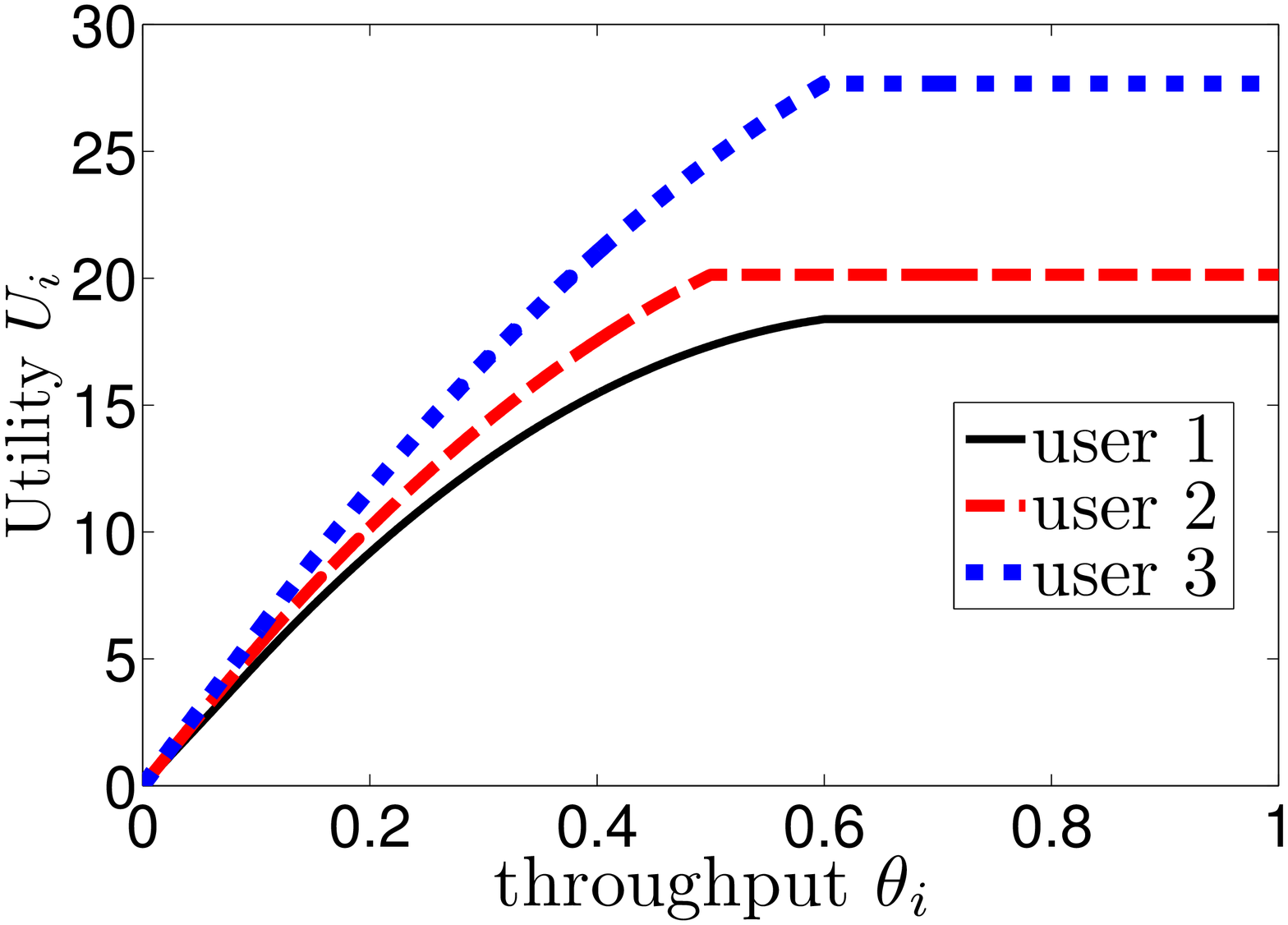}
                \caption{Utility Functions of the Three CSMA Users}
                \label{Utility3}
        \end{subfigure}% 
        \caption{Target Rates and Utility Functions of the CSMA Users}\label{TRandUF}
        %\vspace{-1.5em}
\end{figure}

The utility function is designed to provide differentiated treatment for the links based on their actual demand and willingness to pay for the desired transmission rate. For the example given in Fig. \ref{TargetRate3}, D2D link 1 and link 3 have the same range in their target rates, i.e., $\pi_1=\pi_3$ and $\gamma_1=\gamma_3$, and link 3 has a higher willingness to pay, i.e., $m_3>m_1$. When the price $M$ continually decreases from a large value until zero, link 3 will be admitted into the system first.

From the game theoretic perspective, link $i$ will try to choose its target rate $\tilde{\theta}_i$ in order to maximize its own payoff $v_i(\theta_i)=U_i(\theta_i)-M\theta_i$ (utility minus cost). To be compatible with such an incentive, we can reversely derive the utility function of D2D link $i$ as follows. If the utility function $U_i(\theta_i)$ is concave, then the $\theta_i$ value that maximizes $v_i(\theta_i)$ is given by the first-order condition $v_i'(\theta_i)=0$, i.e., $\tilde{\theta}_i=(U_i')^{-1}(M)$. Equating with the example of (\ref{TargetRate}) and we can reversely derive the utility function for D2D link $i$:
\begin{equation}\label{Utility}
U_i(\theta_i)=\left\{
    \begin{array}{l l l}
      m_i\theta_i,&\textrm{ $0\leq \theta_i < \gamma_i$,}\\
      m_i\theta_i-\frac{(\theta_i-\gamma_i)^2}{2b_i},&\textrm{ $\gamma_i\leq \theta_i< \pi_i$,}\\
      m_i\pi_i-\frac{(\pi_i-\gamma_i)^2}{2b_i},&\textrm{ $\pi_i\leq\theta_i\leq 1$}.\\
    \end{array} \right.
\end{equation}
The utility functions for the three links' example in Fig. \ref{TargetRate3} are plotted in Fig. \ref{Utility3}.
If we take the derivative of $U_i(\theta_i)$ on $\theta_i$ in (\ref{Utility}), it can be seen that a higher $m_i$ value corresponds to a steeper slope, which suggests a higher willingness to pay. This can be seen from Fig. \ref{Utility3}, in which the utility function of link 3 has a steeper slope than that of link 1.

\subsection{A Subgame of Noncooperative CSMA Users}\label{CSMAgame}

The Stackelberg game at the $l$-th stage begins with the BS broadcasting a price $M^{(l)}$. Each D2D link $i$ ($i \in \mathcal{N}$) aims to maximize its payoff $v_i(\theta_i)=U(\theta_i)-M^{(l)}\theta_i$. According to the analysis in Section \ref{UtilityFunction}, link $i$'s objective is equivalent to attaining a target rate $\tilde\theta_i^{(l)}$ under the service price $M^{(l)}$, as given in (\ref{TargetRate}). Whether these target rates are achievable still depends on whether the underlying CSMA mechanism can support these transmissions. The D2D links therefore play a CSMA game among themselves to determine their individual TAs to make the throughput ${\theta_i}$ as close to $\tilde\theta_i^{(l)}$ as possible. This CSMA game is therefore the follower-subgame in the Stackelberg game. 
To avoid the links from transmitting too aggressively and driving the network to unstable states as a result of congestion, a simple approach is to let $M$ begin with a large value and then gradually decrease. %The D2D links will therefore avoid from ended at a congested situation. 

We formally state the CSMA subgame as follows:

\textit{Players}: Distributed Tx-Rx pairs (D2D links), $i\in \mathcal{N}$, who compete to transmit in the ideal CSMA network.

\textit{Strategies}: Each player $i$ chooses its TA $r_i\in \mathcal{R}$, $\forall i \in \mathcal{N}$.

\textit{Objectives}: Each player $i$ ($i \in \mathcal{N}$) aims to achieve its target rate $\tilde\theta_i^{(l)}$ set by maximizing its payoff 
\begin{equation}\label{payoff}
	v_i(\theta_i)=U(\theta_i)-M^{(l)}\theta_i
\end{equation}
under the given service price $M^{(l)}$. 
%According to the analysis in Section \ref{UtilityFunction}, player $i$'s objective is equivalent to attaining a target rate under the service price $M$, as given in (\ref{TargetRate}).

Note that the throughput $\theta_i$ that player $i$ can achieve is determined by its own TA $r_i$ and the TAs of all the other players $r_{-i}$ based on the relationships in (\ref{throughput}). The equilibrium solution of the CSMA subgame is a \textit{Nash Equilibrium (NE)}\cite{Fudenberg}, which is defined as a strategy profile $\vec{r}^*=[r_1^*,\cdots,r_N^*]$ in which player $i$'s strategy $r_i^*$ is a best response to the strategies $r_{-i}^*$ of all the other players, i.e., 
\begin{equation}\label{NashEquilibrium}
r_i^*= \arg \min_{r_i\in (-\infty,+\infty)} |\tilde{\theta}_i^{(l)}-{\theta}_i(r_i,r_{-i}^*)|, \forall i \in \mathcal{N},
\end{equation}%
where $\theta_i(\vec{r})$ is the achieved throughput of D2D link $i$ in the ICN model, as given in (\ref{throughput}).

It is clear that the objective of the subgame is to find the equilibrium TAs for all D2D links so that every link achieves throughputs that are as close to what are desired. 
According to Lemma \ref{CSMAlemma}, if the target rate ${\vec{\tilde\theta}}$ is in $\mathcal{C}$, i.e., it is achievable, then there exists a unique TA $\vec{r}^*$ such that $\theta_i^*(\vec{r}^*) =\tilde\theta_i^{(l)}, \forall i\in\mathcal{N}$. On the other hand, if the target rate ${\vec{\tilde\theta}}$ is beyond $\mathcal{C}$, during the myopic best response updates, all players will keep increasing their TAs if their throughputs are lower than their respective target rates, i.e., all links transmit aggressively and result in undesired network congestion.
The existence and uniqueness of the NE in the CSMA subgame can then be established in the following proposition.

\newtheorem{proposition}{Proposition}
\begin{proposition}\label{csmaNEexistence}
For the target rate $\vec{\tilde{\theta}}\in \mathcal{C}$ (strictly feasible region), there exists a unique finite-valued NE $\vec{r}^*\in \mathcal{R}^N$ in the CSMA subgame.  Moreover, the target rate $\vec{\tilde{\theta}}$ is achieved at the NE, i.e., $\vec{\theta}^*(\vec{r}^*)=\vec{\tilde{\theta}}$.
\end{proposition}

\textit{Proof}: 
From Lemma \ref{CSMAlemma}, if the target rate $\vec{\tilde{\theta}}\in \mathcal{C}$ (strictly feasible region), then there exists a unique finite-valued $\vec{r}^*\in \mathcal{R}^N$ such that $\vec{\theta}^*(\vec{r}^*)=\vec{\tilde{\theta}}$.
As can be seen from (\ref{NashEquilibrium}), there exists a unique NE $\vec{r}^*$, since the payoff of each player is maximized when $\vec{\theta}^*(\vec{r}^*)=\vec{\tilde{\theta}}$ and no player has the incentive to deviate from this NE unilaterally.
$\blacksquare$

In practice, the strategies of all the other players $r_{-i}$ are usually not known by player $i$ if we assume that there is no explicit information exchange among the players. We therefore design some distributed updating method for the players to arrive at the NE. Let each player update its strategy by measuring its own local statistics, e.g., measured throughput $\hat{\theta}_i$. For the $k$-th measurement period $\tau(k)$, player $i$ keeps a record of the accumulated transmission time, $T_i(k)$, and obtains the empirical average throughput as
\begin{equation}\label{MessuredTheta}
\hat{\theta}_i(k)=T_i(k)/\tau(k), \forall i \in \mathcal{N}.
\end{equation} %
A distributed way for player $i$ to update its strategy can be
\begin{equation}\label{IterationDynamics}
r_i(k+1)=r_i(k)+\alpha\cdot(\tilde{\theta}_i^{(l)}-\hat{\theta}_i(k)), \forall i \in \mathcal{N},
\end{equation} %
where $\alpha$ is a small positive step size.
The conditions for the convergence of the NE in the CSMA subgame are summarized in the following proposition.

\begin{proposition}\label{csmaNEconvergence}
For the target rate $\vec{\tilde{\theta}}\in \mathcal{C}$ (strictly feasible region), the iteration dynamics in (\ref{IterationDynamics}) with a small enough step size $\alpha$ and a long enough measurement period $\tau$ will always converge to the NE $\vec{r}^*\in \mathcal{R}^N$ in the CSMA subgame. 
\end{proposition}

\textit{Proof}: 
We apply the same idea used when proving Lemma \ref{CSMAlemma}. Given a $\vec{\tilde{\theta}}\in \mathcal{C}$, we use the maximum log-likelihood method to estimate the parameters $\vec{r}^*$ which result in $\vec{\theta}(\vec{r}^*)=\vec{\tilde{\theta}}$, or equivalently, result in the desired state probability distribution $\overrightarrow{p}^{\tilde{\theta}}$ such that
$\vec{\theta}(\vec{r}^*)=\textstyle\sum_{\vec{s}\in\mathcal{S}}p_{\vec{s}}^{\tilde{\theta}}\vec{s}$.
The log-likelihood function $F(\vec{r};\vec{\tilde{\theta}})$ is given in (\ref{LogLikelihood}).
It has been shown in Section \ref{SectionLemma} that $F(\vec{r};\vec{\tilde{\theta}})$ is a strictly concave function in $\vec{r}$ and attains its maximum when $\vec{\theta}(\vec{r}^*)=\vec{\tilde{\theta}}$.
Therefore, we can use the subgradient method \cite{boyd2003subgradient} to obtain the optimal solution $\vec{r}^*$,
where $\tilde{\theta}_i^{(l)}-\hat{\theta}_i(k)$ is an estimation of the gradient $\frac{\partial F(\vec{r};\vec{\tilde{\theta}})}{\partial r_i}$ (see (\ref{FirtOrderPartial})) in the $k$-th measurement period.
Since the objective function $F(\vec{r};\vec{\tilde{\theta}})$ is differentiable and concave in $\vec{r}$, the subgradient method with constant step size $\alpha$ yields convergence to the optimal value, provided the step size $\alpha$ is small enough and the measurement period is long enough \cite{boyd2003subgradient}.
In summary, the above proposition follows.
$\blacksquare$

As mentioned above, under the myopic best response update approach, if the target rate ${\vec{\tilde\theta}}$ is beyond $\mathcal{C}$, all players will keep increasing their TAs if their throughputs are lower than their respective target rates, and the network will be pushed into an undesired congested situation. To overcome this problem,   
%Since $r_i=\log\rho_i$ and $\rho_i=E[t_{tr,i}]/E[t_{cd,i}]$, if we assume that the mean transmission time $E[t_{tr,i}]$ is finite, then $r_i\rightarrow +\infty$ requires an infinitely small mean countdown time $E[t_{cd,i}]$, which is usually not practical in the CSMA method.
we impose a upper limit $r_{max}$ on $r_i, \forall i\in \mathcal{N}$ as an implementation constraint.
The iteration dynamics in (\ref{IterationDynamics}) then become:
\begin{equation}\label{IterationDynamicsRmax}
r_i(k+1)=\min\{ r_i(k)+\alpha\cdot(\tilde{\theta}_i^{(l)}-\hat{\theta}_i(k)), r_{max}\}, \forall i \in \mathcal{N}.
\end{equation} %
The physical meaning of imposing the $r_{max}$ constraint is to refrain the D2D links from transmitting too aggressively, so that the local congestion at some links will not affect the whole network.
The outcome of introducing such a restriction is that the feasible throughput region will shrink into a subset of the original one. Hence the solution obtained with this constraint imposed is always ensured to be within $\mathcal{C}$. During the myopic play, if any link arrives at $r_{max}$, the BS will be informed. Then the price $M$ is frozen and the whole D2D network functions at the boundary of the ``shrunken" feasible throughput region. If this happens, not all users are able to achieve their desired rates or even admitted, as the network is in ``congestion".

\subsection{Analysis of the Stackelberg Game}\label{QuasiConvex}
In this subsection we analyze the game structure of the Stackelberg game. % as a basis for designing the pricing strategies for the leader.
From Proposition \ref{csmaNEexistence} and Proposition \ref{csmaNEconvergence}, to satisfy $\vec{\tilde{\theta}}\in\mathcal{C}$, it is equivalent to checking that the CSMA subgame converges to the unique subgame NE $\vec{r}^*$ defined in (\ref{NashEquilibrium}) under the given service price $M$.
Therefore, the leader problem in (\ref{LeaderProblemC}) is equivalent to the following optimization problem:
\begin{equation}\label{LeaderProblem}
\begin{array}{l l}
\max\limits_{M\geq 0} ~~g(M)\\
\mbox{s.t.} ~~\left\{
    \begin{array}{l l l}
      \textrm{equality constraint (\ref{TargetRate})}, \forall i\in \mathcal{N}, \\ 
      \textrm{equality constraint (\ref{NashEquilibrium})},\\
      r_i^*< r_{max}, \forall i\in \mathcal{N},\\
    \end{array} \right.
\end{array}
\end{equation}%
where $r_i^*$ is the TA of D2D link $i$ at the NE of the CSMA subgame, as given in (\ref{NashEquilibrium}).
The constraint $r_i^*< r_{max}, \forall i\in \mathcal{N}$ implies that the target rate $\vec{\tilde{\theta}}$ as given in (\ref{TargetRate}) is strictly feasible at the subgame NE, i.e., the throughput will converge to $\vec{\theta}^*(\vec{r}^*)=\vec{\tilde{\theta}}$ in the CSMA subgame.

To distinguish from the NE $\vec{r}^*$ in the CSMA subgame, we call the equilibrium solution $(M^{opt},\vec{r}^{opt})$ of the Stackelberg game as the Stackelberg Equilibrium (SE) \cite{StackelbergEquilibrium}, where $M^{opt}$ is the optimal solution to (\ref{LeaderProblem}) and $\vec{r}^{opt}$ is the subgame NE under the price $M^{opt}$.

The problem in (\ref{LeaderProblem}) is non-convex \cite[pp. 136]{Boyd} since the objective function $g(M)=\textstyle\sum_{i=1}^N \tilde{\theta}_i(M)$ is non-concave in $M$ and the equality constraints (\ref{TargetRate}) and (\ref{NashEquilibrium}) are nonlinear.
Fortunately, the problem can be converted into a quasi-convex optimization problem \cite[pp. 144]{Boyd} and the solution can be iteratively evaluated by solving a sequence of convex optimization problems. It can be interpreted in the following way.
Since the target rate $\tilde{\theta}_i(M)$ of each D2D link $i$ is non-increasing with the price $M$, 
the chain of prices $M^{(0)}>M^{(1)}>\cdots>M^{(l)}>M^{(l+1)}$ induces a chain of target rates $\vec{\tilde{\theta}}^{(0)}\preceq\vec{\tilde{\theta}}^{(1)}\preceq\cdots \preceq\vec{\tilde{\theta}}^{(l)}\preceq\vec{\tilde{\theta}}^{(l+1)}$.
Therefore, the objective function $g(M)=\textstyle\sum_{i=1}^N \tilde{\theta}_i(M)$ is also non-increasing with $M$, and hence is quasi-concave in $M$.
Regarding the constraints in (\ref{LeaderProblem}),
from Lemma \ref{CSMAlemma}, if the target rate $\vec{\tilde{\theta}}^{(l)}\in \mathcal{C}$ (strictly feasible throughput region, which is the interior of the feasible throughput region $\bar{\mathcal{C}}$), then it is achievable with finite-valued TAs $\vec{r}^*$. 
On the other hand, from Theorem \ref{PartialOrderTheorem}, if the target rate $\vec{\tilde{\theta}}^{(l)}\not\in \bar{\mathcal{C}}$, then any target rate $\vec{\tilde{\theta}}^{(l+1)}\succeq\vec{\tilde{\theta}}^{(l)}$ is not in $\bar{\mathcal{C}}$, i.e., it is not achievable and the constraints in (\ref{LeaderProblem}) are not satisfied. 
The crossing from within $\bar{\mathcal{C}}$ to beyond can be detected by the use of $r_{max}$. 
Therefore, the superlevel set $\{M|g(M)\geq G\}$ is convex, which is equivalent to the line segment $\{M|g^{-1}(\sup g)\leq M\leq g^{-1}(G)\}$, where $G$ is a constant, $g^{-1}$ is the inverse function of $g(M)$, and $\sup g$ is the optimal value of (\ref{LeaderProblem}).

In summary, the problem in (\ref{LeaderProblem}) is quasi-convex \cite[pp. 137, pp.144]{Boyd}, since the objective function $g(M)$ to be maximized is quasi-concave, and the superlevel set $\{M|g(M)\geq G\}$ is convex.
As a result, the problem in (\ref{LeaderProblem}) can be reduced into a sequence of feasibility problems:
\begin{equation}\label{FeasibilityProb}
\begin{array}{l l}
\textrm{find} ~~M\\
\mbox{s.t.} ~~\left\{
    \begin{array}{l l l l}
      g(M)\geq G,\\
      \textrm{equality constraint (\ref{TargetRate})}, \forall i\in \mathcal{N},\\
      \textrm{equality constraint (\ref{NashEquilibrium})},\\
      r_i^*< r_{max}, \forall i\in \mathcal{N}.\\
    \end{array} \right.
\end{array}
\end{equation}%
If the problem (\ref{FeasibilityProb}) is feasible, then the maximum total throughput $\sup g$ is not less than $G$. Conversely, if the problem (\ref{FeasibilityProb}) is infeasible, then we can conclude $\sup g<G$.
In order to find the optimal value $\sup g$ to the problem (\ref{LeaderProblem}), we can test different superlevels $G$ in the feasibility problem (\ref{FeasibilityProb}). 
%The simple bisection method \cite[pp. 145]{Boyd} can be used to narrow down the range of $G$ and converge to the optimal value $\sup g$.

For each superlevel $G$,
from the proof of Proposition \ref{csmaNEconvergence}, the feasibility problem in (\ref{FeasibilityProb}) is equivalent to the following max-log-likelihood problem:
% which estimates the parameters $\vec{r}$ that achieves the target rate $\vec{\tilde{\theta}}$ under the given price $M=g^{-1}(G)$:
\begin{equation}\label{MaxLog}
\begin{array}{l l}
\max\limits_{\vec{r}} ~~F(\vec{r};\vec{\tilde{\theta}})\\
\mbox{s.t.} ~~\left\{
    \begin{array}{l l l}      
      \textrm{equality constraint (\ref{TargetRate})}, \forall i\in \mathcal{N},\\  
      M=g^{-1}(G),\\
      r_i< r_{max}, \forall i\in \mathcal{N},\\
    \end{array} \right.
\end{array}
\end{equation}
where $F(\vec{r};\vec{\tilde{\theta}})$ is the log-likelihood function defined in (\ref{LogLikelihood}).
In other words, if the problem (\ref{FeasibilityProb}) is feasible,
then there exists a price $M=g^{-1}(G)$, such that the target rate $\vec{\tilde{\theta}}(M)$ is achievable with finite TA $r_i< r_{max}, \forall i\in \mathcal{N}$.
Therefore, we can use the max-log-likelihood method to estimate the parameters $\vec{r}$ which achieve the target rate $\vec{\tilde{\theta}}(M)|_{M=g^{-1}(G)}$, as given in (\ref{TargetRate}).
Notice that given the constant $G$, the price $M$ and the target rate $\vec{\tilde{\theta}}$ become constant values as well.
%Therefore, the constraints in (\ref{MaxLog}) are all linear.
Moreover, as shown in the proof of Proposition \ref{csmaNEconvergence}, the log-likelihood function is concave in $\vec{r}$.
As a result, the max-log-likelihood problem in (\ref{MaxLog}) is a convex optimization problem, and can be solved by the subgradient updating method in (\ref{IterationDynamicsRmax}).

If the iteration dynamics converge to a subgame NE with $r_i^*< r_{max},\forall i\in\mathcal{N}$, then the optimal solution to (\ref{MaxLog}) exists, i.e., the problem (\ref{FeasibilityProb}) is feasible. Otherwise, if the iteration dynamics in (\ref{IterationDynamicsRmax}) converge to a subgame NE with $r_i^*= r_{max}$ and $\theta_i^*<\tilde{\theta}_i$ for some D2D link $i$, then the optimal solution to (\ref{MaxLog}) does not exist and the problem (\ref{FeasibilityProb}) is infeasible, i.e., not all D2D links' target rates are being achieved.

In summary, the problem in (\ref{LeaderProblem}) can be reduced into a sequence of convex optimization problems. A simple bisection method can be used to choose the superlevels $G$ (or equivalently, the price $M=g^{-1}(G)$) and test the feasibility problem (\ref{FeasibilityProb}). Alternatively, we can borrow ideas from the feasible direction method \cite[Chap. 10]{Bazaraa} which avoids testing in the infeasible region, and design the pricing strategies so as to keep the network operating in the feasible region while tuning the price $M$. 

\subsection{Pricing Strategies of the Stackelberg Leader}\label{PricingStrategies}
We call each round of CSMA subgame under a certain price $M$ as a \textit{stage} in the Stackelberg game. In each stage, the leader needs the feedback from each D2D link $i$ about its target rate $\tilde{\theta}_i$, and 
the converged TA $r_i^*$ and throughput $\theta_i^*$.
Notice that the leader only has knowledge about the monotonicity of the D2D link's target rate with the price $M$ and no information about (\ref{TargetRate}) of all links is required.

Under some low load situations, all links achieve their maximum desired throughput $\pi_i, \forall i\in \mathcal{N}$, if for all links $i\in\mathcal{N}$, the target rate $\tilde{\theta}_i>0$ and remains unchanged between two consecutive prices $M^{(l)}$ and $M^{(l+1)}$.
Under heavy load situations, the pricing strategies of the Stackelberg leader need to be carefully designed to converge to the optimal price $M^{opt}$.

To detect convergence, we 
define $\Delta_i=r_{max}-r_i^*$ as the ``margin" of transmission aggressiveness for each D2D link $i\in \mathcal{N}$. When the achieved target rates are close to the capacity boundary, the leader can make use of $\Delta_{min}=\min \{\Delta_i,\forall i\in \mathcal{N}\}$ as an indication of how close the current throughput $\vec{\theta}^*$ is to the boundary of $\mathcal{C}$.
%constrainted Pareto front $\mathcal{P}_c$. 
%In particular, $\Delta_{min}=0$ indicates that $\vec{\theta}^*\in\mathcal{P}_c$.
Since the total throughput $g(M)$ is non-increasing with the price $M$, the leader can gradually decrease $M$ to increase $g(M)$ until the constraint $r_i^*< r_{max}$ is ``critically" satisfied for some D2D link says $i$, i.e., $\Delta_{min}\leq\epsilon$, where $\epsilon$ is a small positive threshold.

The algorithm at the BS works as follows.
In the 0-th stage, the leader can start with a large price $M^{(0)}$ so that the network starts with low load. 
Similar to the Newton method \cite[pp. 488]{Boyd} which applies line search to narrow down the searching region before using Newtonian steps to refine the optimal solution, the adjustment of our price strategies consist of two phases as well. In the first phase the leader uses a relatively large decrement step $\phi$ to decrease price $M$ until $\Delta_{min}\leq\eta$, where $\eta>\epsilon$ is a threshold before entering the second phase. In the second phase, the decrement steps are refined using $\Delta_{min}$ since $\Delta_{min}$ is getting smaller as the target rates are approaching the boundary of the feasible throughput region.

In summary, the leader can update its price $M$ based on $\Delta_{min}$ at the end of the $l$-th stage as follows:
%\begin{equation}\label{PriceUpdate}
%M^{(l+1)}=\max\{ M^{(l)}-\beta\cdot\Delta_{min}, M_{lower}\},
%\end{equation} %
\begin{equation}\label{PriceUpdate}
\begin{array}{l l}
M^{(l+1)}=\left\{
    \begin{array}{l l}
      M^{(l)}-\phi, &\textrm{ $\Delta_{min}>\eta$,}\\
      \max\{ M^{(l)}-\beta\cdot\Delta_{min}, M_{lower}\}, & \textrm{ $\Delta_{min}\leq\eta$},\\
    \end{array} \right.
\end{array}
\end{equation}
where $\phi$ is a positive constant, $\beta$ is a positive parameter. $M_{lower}$ is initially set at 0 and is updated to take the value of current $M^{(l)}$ once it is detected that the solution for the target rate $\vec{\tilde{\theta}}$ is outside the feasible region. Its purpose is to ensure that subsequent $M^{(l+1)}, \cdots$ should not go below this value.
%Such a pricing method belongs to the broad class of feasible direction methods, where $\Delta_{min}>0$ suggests that decreasing $M$ is the feasible direction, and the amount of decrement is being refined by $\Delta_{min}$ since $\Delta_{min}$ is getting smaller as the target rates are approaching the boundary of $\mathcal{C}$.
The parameter $\beta$ can be chosen to be small enough so that the price gradually decreases until $\Delta_{min}\leq\epsilon$.
However, for faster convergence, it might happen that the initially chosen $\beta$ is too large such that the new price $M^{(l+1)}$ pushes the target rate $\vec{\tilde{\theta}}$ to be outside the feasible region, i.e., $r_i^*=r_{max}$ but $\theta_i^*<\tilde{\theta}_i$ for some D2D link $i$.
In such cases, the leader stores the current unachievable price as the new lower bound $M_{lower}$, resets the price to the previously found achievable price $M_{prev}$, and reduces $\beta$ by a discount factor $\sigma$, e.g., $\sigma=0.9$.

The pricing strategies of the leader and the CSMA subgame are summarized in Algorithm \ref{IterationProcessStackelberg}.
Through Algorithm \ref{IterationProcessStackelberg}, the Stackelberg game is guaranteed to gradually converge to the optimal price $M^{opt}$ under which the total throughput of the CSMA users are maximized while their heterogeneous target rates can all be satisfied.

\begin{algorithm}
\caption{Iteration Process of the Stackelberg Game}\label{IterationProcessStackelberg}
\begin{algorithmic}[1]
\small
\BState \textbf{Initialize}:
\State The BS chooses the initial price $M=M^{(0)}$ and informs the D2D links in the control plane;
\State Each D2D link $i\in \mathcal{N}$ chooses the initial TA $r_i(0)$;

\
\Repeat:
\State In the $l$-th stage:
	\For {$i=1,\cdots,N$ D2D links}:
	\State \begin{varwidth}[t]{0.85\linewidth}
        Set target rate $\tilde{\theta}_i^{(l)}$ based on price $M^{(l)}$, as in (\ref{TargetRate});
      \end{varwidth}	
	\EndFor
	
\State
\Repeat:
\State In the $k$-th measurement period:
	\For {$i=1,\cdots,N$ users}:
	\State \begin{varwidth}[t]{0.8\linewidth}
        Estimate the empirical throughput $\hat{\theta}_i(k)$, as in (\ref{MessuredTheta});
      \end{varwidth}
	\State \begin{varwidth}[t]{0.8\linewidth}
        Update the TA $r_i(k+1)$, as in (\ref{IterationDynamicsRmax});
      \end{varwidth}	     
	\EndFor
	\State $k\leftarrow k+1$;
\Until{
\begin{varwidth}[t]{0.7\linewidth}
        $\vec{r}$ converges to the subgame NE $\vec{r}^*$. % and throughput $\vec{\theta}^*$.
      \end{varwidth}
\State Each user $i\in \mathcal{N}$ informs the BS about $\tilde{\theta}_i^{(l)}$, $r_i^*$ and $\theta_i^*$;     
}

\State
\State At the BS:
\State $\Delta_{min}=\min\limits_{i\in \mathcal{N}} \Delta_i=\min\limits_{i\in \mathcal{N}} (r_{max}-r_i^*)$;

\If {$\Delta_{min}>\epsilon$}:
	\State set $M^{(l+1)}$ as in (\ref{PriceUpdate}); $M_{prev}=M^{(l)}$; $\Delta_{prev}=\Delta_{min}$.
\ElsIf {$0<\Delta_{min}\leq\epsilon$ or ($\vec{\tilde{\theta}}^{(l)}\succ\vec{0}$ and $\vec{\tilde{\theta}}^{(l)}=\vec{\tilde{\theta}}^{(l-1)}$)}:
\State The Stackelberg game converges with $M^{opt}=M^{(l)}$; go to \textbf{END}.
\ElsIf {$r_i^*=r_{max}$ but $\theta_i^*<\tilde{\theta}_i$ for some user $i$}:
\State $M_{lower}=M^{(l)}$; $\beta\leftarrow\sigma\cdot\beta$; $M^{(l+1)}= \max\{ M_{prev}-\beta\cdot\Delta_{prev}, M_{lower}\}$.
\EndIf
	\State $l\leftarrow l+1$;
\Until{The Stackelberg game converges. \textbf{END}.}
\end{algorithmic}
\end{algorithm}
%\vspace{-2.0em}

An important side-information which can be provided by the proposed algorithm is the identification of the bottleneck link in the heterogeneous D2D networks.
Upon convergence of the Stackelberg game, the D2D link $L=\arg\min \{\Delta_i,\forall i\in \mathcal{N}\}=\arg\min \{r_{max}-r_i^{opt},\forall i\in \mathcal{N}\}$ is the bottleneck link to the network since any further decrease on the price $M^{opt}$ would drive the target rate $\vec{\tilde\theta}$ to be outside the capacity region and link $L$ can no longer achieve its target rate. The identification of such bottleneck links can be of valuable information, for example, in data offloading, to re-assign these links back to the cellular network when necessary. 
%The data is either transmitted through the D2D link or routed through the BS using cellular links.
%A possible way to improve system performance is to remove the bottleneck link $L$ in the D2D network and port link $L$'s traffic towards the conventional cellular mode. 
How to achieve optimal trade-off remains as interesting future work.

\subsection{Complexity of Algorithm \ref{IterationProcessStackelberg}}\label{SectionComplexity}

Algorithm \ref{IterationProcessStackelberg} consists of two loops. In the outer loop, the BS chooses a service price $M^{(l)}$ at the $l$-th stage according to the pricing strategies in Section \ref{PricingStrategies}. In the inner loop, for each given service price $M^{(l)}$, the D2D links play the CSMA subgame distributively and iteratively until converging to their respective target rates. We analyze the complexity in terms of the number of iterations required, first for the CSMA subgame, then for the pricing strategies.

For the CSMA subgame, assume that the target rate $\vec{\tilde\theta}$ under the given service price $M^{(l)}$ is in the strictly feasible region $\mathcal{C}$. According to Proposition \ref{csmaNEexistence} and Proposition \ref{csmaNEconvergence}, the distributed strategy updates of the CSMA users in (\ref{IterationDynamics}) are equivalent to the gradient method in maximizing the log-likelihood function $F(\vec{r};\vec{\tilde{\theta}})$ which is differentiable and strictly concave in $\vec{r}$.
In particular, the gradient of $F(\vec{r};\vec{\tilde{\theta}})$ is $\nabla F(\vec{r};\vec{\tilde{\theta}})=\vec{\tilde\theta}-\vec{\theta}(\vec{r})$, as shown in (\ref{FirtOrderPartial}). Since the maximum value of $F(\vec{r};\vec{\tilde{\theta}})$ is finite and attained at $\vec{r}^*$, this means that $\vec{\theta}^*(\vec{r}^*)=\vec{\tilde\theta}$ can be solved by setting the gradient $\nabla F(\vec{r}^*;\vec{\tilde{\theta}})=\vec{0}$.

Since the norms of the throughput $\vec{\theta}(\vec{r})$ and its gradient $\nabla \vec{\theta}(\vec{r})$ are both bounded, it can be shown that $\nabla F(\vec{r};\vec{\tilde{\theta}})$ is Lipschitz continuous \cite{polyak1987} in $\vec{r}$, i.e., $\|\nabla F(\vec{r}_a;\vec{\tilde{\theta}})-\nabla F(\vec{r}_b;\vec{\tilde{\theta}})\|=\|\vec{\theta}(\vec{r}_a)-\vec{\theta}(\vec{r}_b)\|\leq H\|\vec{r}_a-\vec{r}_b \|$, $\forall \vec{r}_a,\vec{r}_b\in \mathcal{R}^N$, where $H$ is a positive constant.
According to Theorem 1 in \cite[Section 1.4]{polyak1987} and Theorem 2.1.14 in \cite[Section 2.1.5]{nesterov2004}, for a small enough step size $\alpha$ ($0<\alpha\leq 1/H$), the number of iterations to reach
$\|\nabla F(\vec{r};\vec{\tilde{\theta}})\|=\|\vec{\tilde\theta}-\vec{\theta}(\vec{r})\|<\xi$ is $O(1/\xi)$ (i.e., no more than a fixed multiple of $1/\xi$).

It is worth to mention that although this complexity $O(1/\xi)$ on the number of required iterations is independent on the number of users, we have inherently assumed that the measurement period $\tau$ is long enough to provide an accurate estimation of throughputs. In fact, the choice of $\tau$ depends on the number of users and the underlying topology. The purpose of choosing a large $\tau$ is to ensure that the Markov chain corresponding to the updated $\vec{r}$ reaches its stationary distribution to allow for an accurate estimation of throughputs. In general, a larger number of users requires a larger value of $\tau$. More comparisons and discussions on how to choose $\tau$ for a given number of users and different topologies can be found in \cite{ICNuniqueProof}.

%Note that this upper bound $O(1/\xi)$ on the number of required iterations is independent on the number of users, given that the measurement period $\tau$ is long enough to provide an accurate estimation of throughputs. On the other hand, the choice of $\tau$ is dependent on the number of users and the underlying topology. The purpose of choosing a large $\tau$ is to ensure that %the Markov chain corresponding to 
%the updated $\vec{r}$ reaches its stationary state to allow for an accurate estimation of throughputs. 
%In general, a larger number of users requires a larger value of $\tau$. See \cite{ICNuniqueProof} for more comparisons and discussions on how to choose $\tau$ for a given number of users and different topologies.

We now briefly discuss how to estimate the number of pricing stages required in the outer loop. 
This analysis is complicated by the fact that the step sizes for $M$ are changing each time. Assume that the maximum value of the price is $M_{max}$. In phase 1 of the price setting, since the price is decreasing at a large constant step $\phi$, the number of pricing stages in phase 1 is capped by $\lceil M_{max}/\phi \rceil$, or $\lceil M_{max}/\phi \rceil/2$ on average.
%Assume that the maximum value of the price is $M_{max}$. In phase 1 of the pricing strategies, the price is decreasing at a large constant step $\phi$. Therefore, the maximum number of stages in phase 1 is $\lceil M_{max}/\phi \rceil$.
In phase 2, the TA margin $\Delta_{min}\leq\eta$, and the price is already close to the optimal. In the algorithm, we refine the price change $\delta M^{(l+1)}=M^{(l+1)}-M^{(l)}$ at stage $l$ according to $\delta M^{(l+1)}=-\beta \Delta_{min}^{(l)}$ progressively until the TA margin gradually approaches the required precision $\epsilon$, i.e., $\Delta_{min}\leq\epsilon$.
Assume that the interval $\epsilon<\Delta_{min}\leq\eta$ is small, through simulations we find that the relationship between the TA margin $\Delta_{min}^{(l+1)}$ and the price change $\delta M^{(l+1)}$ can be approximated by $\Delta_{min}^{(l+1)}=\Delta_{min}^{(l)}+B\cdot \delta M^{(l+1)}$, where $B$ is a positive constant. 
Since we set $\delta M^{(l+1)}=-\beta \Delta_{min}^{(l)}$, hence $\Delta_{min}^{(l+1)}=\Delta_{min}^{(l)}+B\cdot (-\beta \Delta_{min}^{(l)})=(1-\beta B)\Delta_{min}^{(l)}$.
The TA margin then follows a geometric progression and we can estimate the value of $h$, so that $\delta M^{(l+h)}\leq \epsilon$.
Hence it can be easily shown that
the number of stages for $\Delta_{min}$ to decrease from $\eta$ to $\epsilon$ is approximately $\frac{\log_{10}\eta/\epsilon}{\log_{10} 1/(1-\beta B)}$, or $O(d\log_{10} (\eta/\epsilon))$ for some suitable choices of $\beta$ and $B$ ($0<\beta B<1$), where $d=\frac{1}{\log_{10} 1/(1-\beta B)}$.
Note that $\beta$ can be chosen according to the value of $B$, but $B$ is topology and utility dependent.
As a result, the total number of stages required for convergence is $O(1/\phi)+O(d\log_{10} (\eta/\epsilon))$.

%The value of $B$ depends on the topology and the utilities of the users, which is usually not known a priori. Fortunately, we can still choose the step size $\beta$ to be small enough so that $0<1-\beta B<1$, and thus $\Delta_{min}$ gradually decreases.
%The number of stages for $\Delta_{min}$ to decrease from $\eta$ to $\epsilon$ is approximately $\frac{\log_{10}\eta/\epsilon}{\log_{10} 1/(1-\beta B)}$.
%Therefore, the total number of stages required for convergence is $O(1/\phi)+O(\log_{10} (1/\epsilon))$.

In summary, the number of iterations required for convergence in the proposed game is given by the number of iterations per stage multiplied by the required number of stages, i.e., $O(1/\xi)\cdot(O(1/\phi)+O(d\log_{10} (\eta/\epsilon)))$.

\section{Simulation Study}\label{Simulation}

\begin{figure}
        \centering
        \begin{subfigure}[b]{0.4\linewidth}
                \includegraphics[width=1\linewidth,  trim=0 0 0 0,clip]{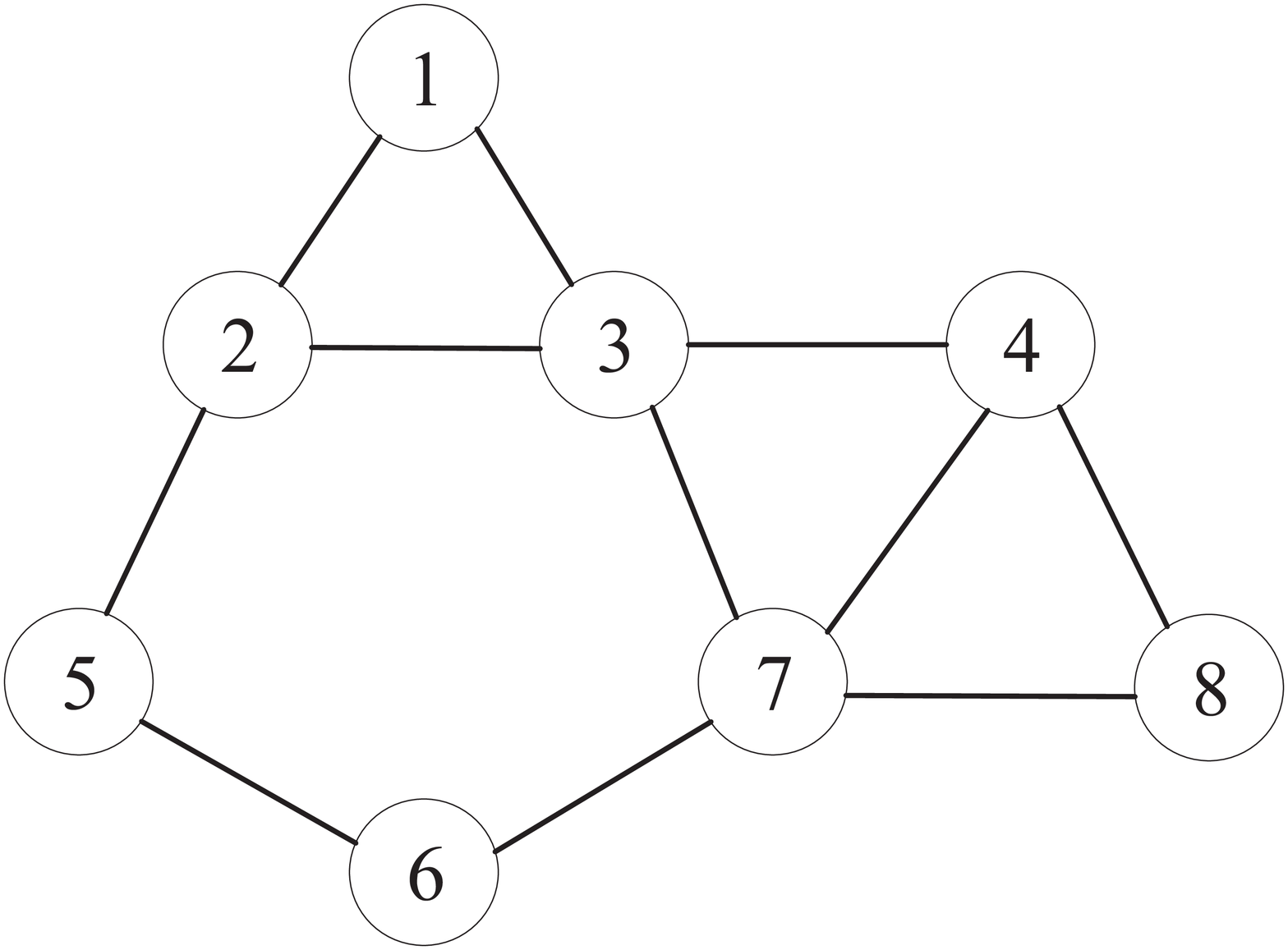}
                \caption{contention Graph}
                \label{asym8}
        \end{subfigure}%
        ~ %add desired spacing between images, e. g. ~, \quad, \qquad, \hfill etc.
        \begin{subfigure}[b]{0.6\linewidth}
                \includegraphics[width=1\linewidth,  trim=0 0 0 0,clip]{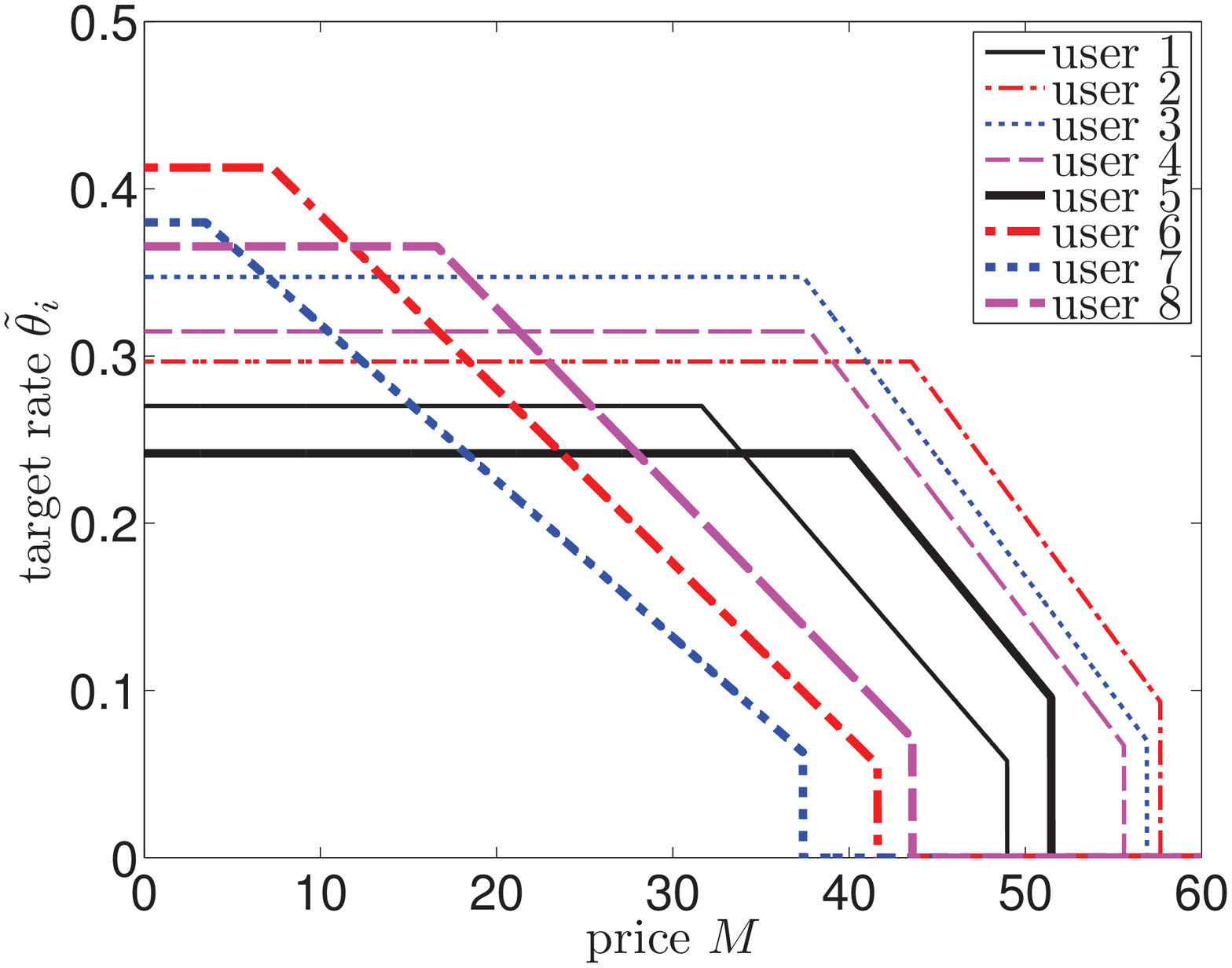}
                \caption{Target Rates under Price $M$}
                \label{utility8}
        \end{subfigure}% 
        \caption{Topology and Target Rates of 8 D2D Links}\label{Stackelberg8}
        %\vspace{-1.5em}
\end{figure}

In this section we demonstrate the Stackelberg game via an example.
Consider the 8 D2D links' contention graph in Fig. \ref{asym8}. 
Assume that the relationships between the links' target rates and the price $M$ are given as in Fig. \ref{utility8}.
In the ICN model, we assume that the links' transmission time is uniformly distributed with mean of 1 ms in the range $[0.5, 1.5]$ ms. Further assume that link $i$'s backoff time is uniformly distributed with mean of $1/\exp(r_i)$ ms in the range $[0,2/\exp(r_i)]$ ms.

\subsection{CSMA Subgame}\label{SimSub}
\begin{figure}[t]
\centering
\includegraphics[width=0.95\linewidth,  trim=0 0 0 0,clip]{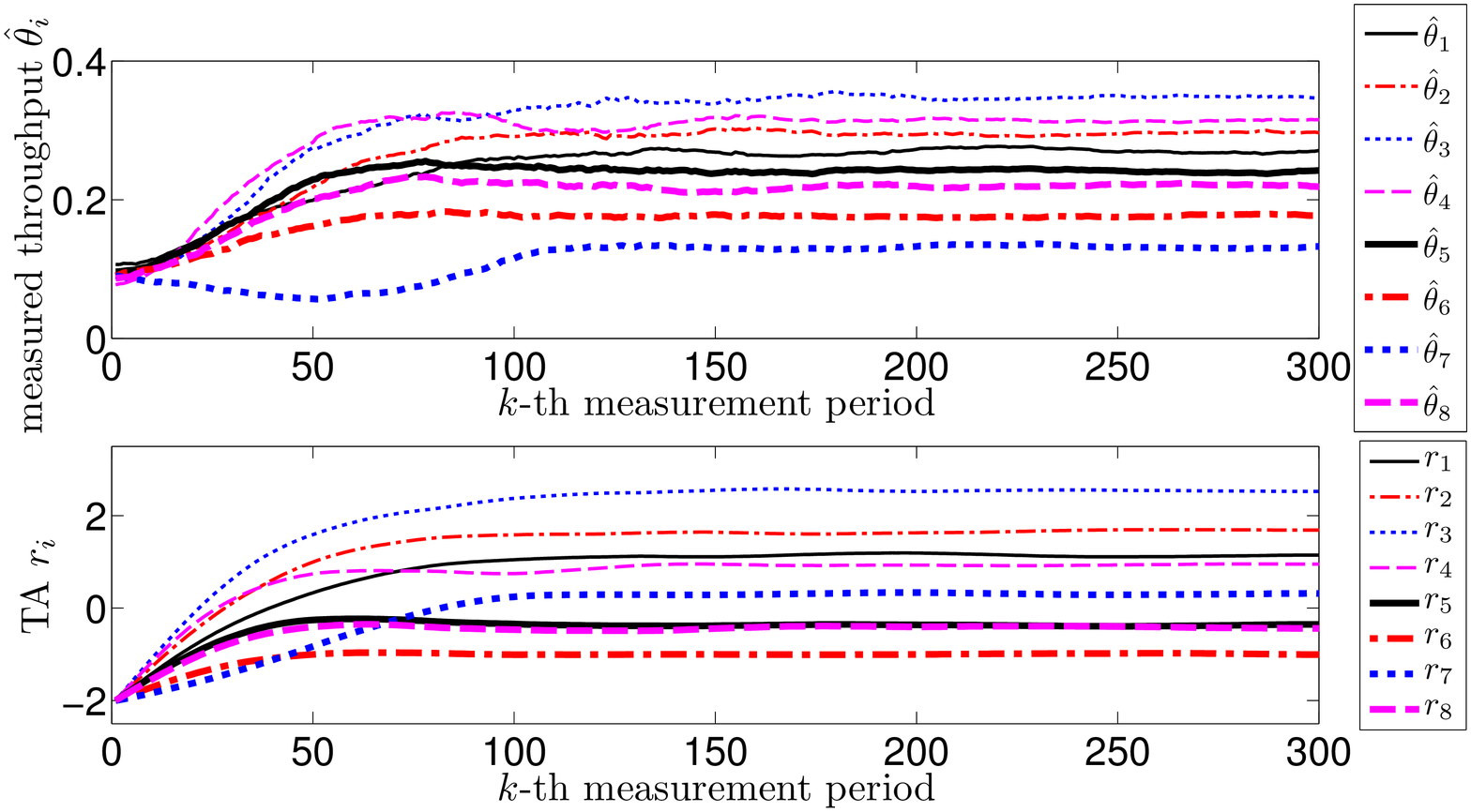} %,height=0.5\linewidth
\caption{CSMA Subgame of the 8 D2D links under $M=30$} \label{resultM}
%\vspace{-1.5em}
\end{figure}

Assume that the current price $M=30$, then from Fig. \ref{utility8} we know that the D2D links' target rates are $\vec{\tilde{\theta}}=[0.270,0.297,0.347,0.315,0.242,0.176,0.132,0.220]$.
Assume that the initial TAs $r_i=-2,\forall i\in\mathcal{N}$. The CSMA subgame is then played according to Lines 10 to 17 in Algorithm \ref{IterationProcessStackelberg}.
In the $k$-th measurement period, we apply a simple averaging filter to smooth the measured throughput as:
\begin{equation}\label{MessuredThetaFactor}
\hat{\theta}_i(k)=(1-\delta)\cdot\hat{\theta}_i(k-1)+\delta\cdot T_i(k)/\tau, \forall i \in \mathcal{N},
\end{equation} %
where $\delta$ is the weight of the new measurement. In our simulations, we choose $\delta=0.05$ and the measurement time $\tau=200$ ms. 
A smaller value of $\delta$ makes the measured throughput more smooth, but also increases the convergence time.
To update TAs as in (\ref{IterationDynamicsRmax}), we choose the step size $\alpha=0.4$ and the maximum allowable TA $r_{max}=3$. 
Note that a smaller value of $\alpha$ guarantees the convergence of the CSMA subgame, but also increases the convergence time.      
The iteration process of the CSMA subgame is then plotted in Fig. \ref{resultM}.
%, in which the curves share the same legend as in Fig. \ref{utility8}.
%As can be seen from Fig. \ref{resultM}, 
The CSMA subgame converges to 99\% of the target rates ($\xi=1\%$) in around 150 iterations. 
%After convergence, the average error of the measured throughputs as compared to the target rates is around $\xi=1\%$. 
According to the complexity analysis in Section \ref{SectionComplexity}, the number of required iterations is $O(1/\xi)$, i.e., in the order of a fixed multiple of $1/\xi=100$. Thus our simulation result is in the same order as the above prediction.

\subsection{Stackelberg Game}
\begin{figure}[t]
\centering
\includegraphics[width=1\linewidth,  trim=50 0 0 0,clip]{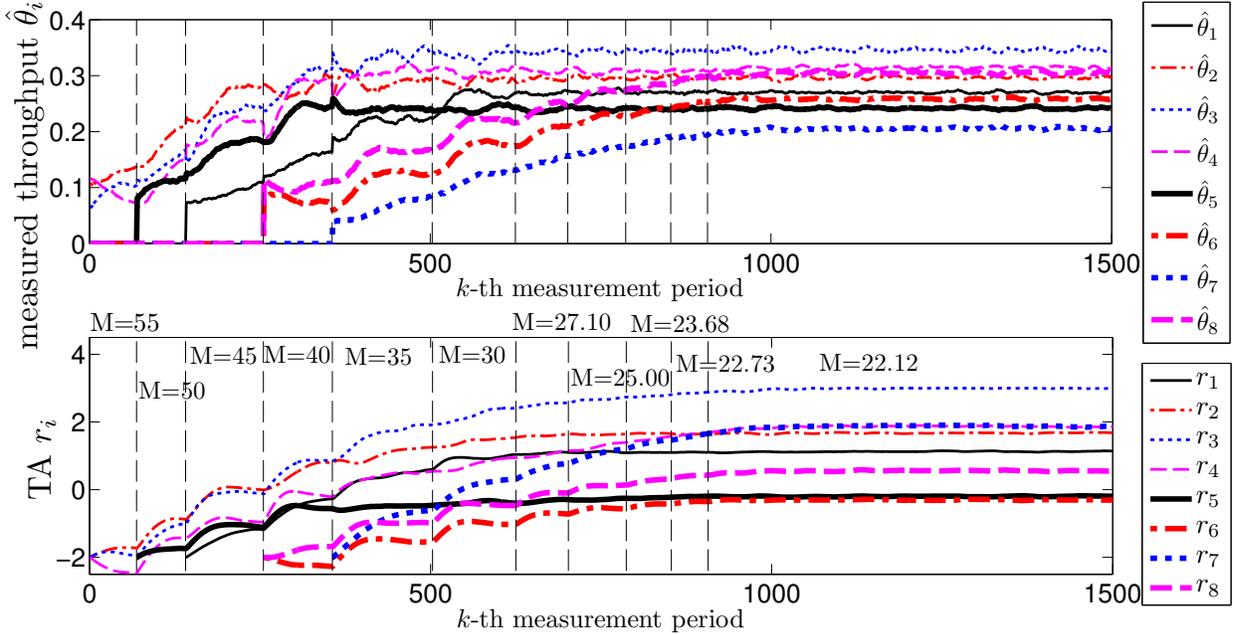} %,height=0.5\linewidth
\caption{Stackelberg Games of the 8 D2D Links} \label{result}
%\vspace{-1.5em}
\end{figure}

Assume that the initial price $M^{(0)}=55$, $
\phi=5$, $\beta=5$, $\eta=1$, $\epsilon=0.1$, $\sigma=0.9$, and the rest of the parameters are the same as in Section \ref{SimSub}.
The iteration process of the Stackelberg game is shown in Fig. \ref{result}.
%, in which the curves share the same legend as in Fig. \ref{utility8}.
The game converges after 11 stages, in which $\Delta_{min}=r_{max}-r_3=3-r_3$ and gradually approaches 0. The first 6 stages undergo a constant price decrement ($\phi=5$), i.e., $M=55,50,45,40,35,30$ until $\Delta_{min}\leq\eta=1$ is detected. After the CSMA subgame converges under the price $M=30$, we have $\Delta_{min}=r_{max}-r_3^*=3-2.4=0.6$ and hence $0.1=\epsilon<\Delta_{min}<\eta$. Therefore, the Stackelberg game enters the second pricing phase, which consists of 5 stages ($M=27.10,25.00,23.68,22.73,22.12$), according to (\ref{PriceUpdate}). We consider the game converged when $\Delta_{min}\approx 0.08<\epsilon$ and the optimal price is $M^{opt}=22.12$.
After convergence, the average error of the measured throughputs as compared to the target rates is around $\xi=1\%$. 
Notice that we cannot decrease $M$ any further since the network is already close to the capacity boundary ($r_3^*=2.92\approx r_{max}=3$). In other words, any further decrease on $M$ would drive the target rate $\vec{\tilde\theta}$ to be outside the capacity region and some D2D links (e.g., link 3) can no longer achieve their target rates.
In summary, the proposed Stackelberg game is able to maximize the total throughput of the CSMA users while the target rates of the heterogeneous users can all be satisfied.

The number of required stages before convergence is consistent with the analysis in Section \ref{SectionComplexity}.
In the above simulations, the maximum value of the price is $M_{max}=60$, and the constant step $\phi=5$.
Therefore, the maximum number of stages in phase 1 is $\lceil M_{max}/\phi \rceil=12$, or 6 on average. In the simulations, phase 1 actually consists of 6 stages before entering phase 2. In phase 2, the required precision $\epsilon=0.1, \eta=1$ and the required number of stages is $O(d\log_{10} (\eta/\epsilon))$, where $d=\frac{1}{\log_{10} 1/(1-\beta B)}$. The value of $B$ can be estimated by using $(\Delta_{min}^{(l+1)}-\Delta_{min}^{(l)})/(M^{(l+1)}-M^{(l)})$, which is approximately 0.07 in the small interval $\epsilon<\Delta_{min}\leq\eta$.
Since we have chosen $\beta=5$, hence $0<\beta B=0.35<1$ and $d=\frac{1}{\log_{10} 1/(1-\beta B)}=5.3$, and the required stages in phase 2 is in the order of a fixed multiple of $d\log_{10} (\eta/\epsilon)=5.3$. In the simulations, phase 1 actually consists of 5 stages before convergence. 
Note that a smaller $\beta$ could be used to guarantee $\beta B<1$, however, it also increases $d$ and hence requires more stages for convergence.
Finally, 
the total number of iterations required in the Stackelberg game is $O(1/\xi)\cdot(O(1/\phi)+O(d\log_{10} (\eta/\epsilon)))$, which is in the order of
$100\cdot(6+5.3)=1130$.
% since the number of iterations in each stage is in the order of a fixed multiple of $1/\xi=100$.
In the simulations, the Stackelberg game actually converges in around 1000 iterations, which is in the same order as the above prediction.

\subsection{Effect of Parameter $r_{max}$}\label{oscilate}

In the above simulations, we have used the parameter $r_{max}=3$, and obtained the optimal price $M^{opt}$ that maximizes the total throughput $g(M)$ for the 8 users in Fig. \ref{Stackelberg8} with heterogeneous rate requirements.
As is discussed at the end of Section \ref{CSMAgame}, by introducing the $r_{max}$ constraint, the feasible throughput region will shrink into a subset of the original one.
In this subsection, we apply different values of $r_{max}$ to the network and obtain the optimal price $M^{opt}$ that maximizes the total throughput $g(M)$ for the 8 users in Fig. \ref{asym8}. To see the effect of parameter $r_{max}$ only, we assume that the 8 users are homogeneous in their rate requirements, i.e., $\gamma_i=0.05, \pi_i=0.55, b_i=0.0125, m_i=50, \forall i\in\mathcal{N}$.

\begin{figure}[t]
\centering
\includegraphics[width=0.95\linewidth,   trim=0 0 0 0,clip]{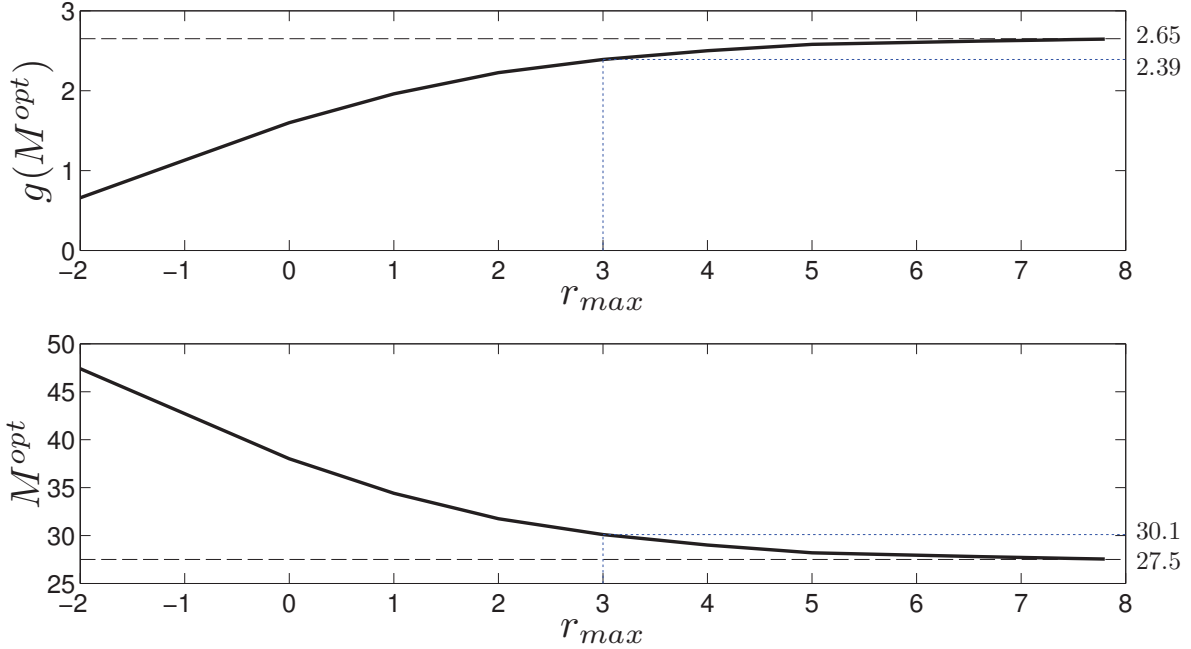} %,height=0.5\linewidth
\caption{Effect of Parameter $r_{max}$} \label{Mrmax}
%\vspace{-1.5em}
\end{figure}

The optimal price $M^{opt}$ and the corresponding total throughput $g(M^{opt})$ under each value of the parameter $r_{max}$ are plotted in Fig. \ref{Mrmax}. From Fig. \ref{Mrmax} we can see that, as the value of $r_{max}$ increases, the achievable total throughput $g(M^{opt})$ also increases, moreover, the rate of increase gradually slows down. 
Recall that $r_{max}$ is used to refrain the D2D links from transmitting
too aggressively.
The outcome of introducing such a restriction is that the feasible throughput region will shrink
into a subset of the original one.
In particular, when $r_{max}=3$, the achievable total throughput is 2.39. For $r_{max}>3$, the total throughput curve becomes almost flat and approaches the upper bound 2.65 when $r_{max}$ tends to infinity (each user achieves a throughput of 0.33), under which the shrunken capacity region stretches back to the original feasible throughput region $\bar{\mathcal{C}}$.
The corresponding bound on the optimal price is $M^{opt}=27.5$. 
In other words, we cannot further 
reduce the price $M$ below 27.5 to increase the target rate $\vec{\tilde{\theta}}$, as it is already on the boundary of the feasible throughput region $\bar{\mathcal{C}}$.

It is observed from Fig. \ref{Mrmax} that a larger $r_{max}$ value leads to a larger capacity region, but this also allows for longer transmission durations.
However, the transmission duration should not be too long in practice, otherwise it would lead to large access delay (where the access delay refers to the time between the onset of two consecutive successful transmissions of a link) and large variations of the delay.
The readers are referred to \cite[Sec. IV]{JiangCSMAcollision} for more discussions.
As a result of the above observations, we have adopted $r_{max}=3$ in the above two subsections. 

\subsection{Unstable Network Behavior without Price Control from BS}

\begin{figure}[t]
\centering
\includegraphics[width=0.95\linewidth,   trim=0 0 0 0,clip]{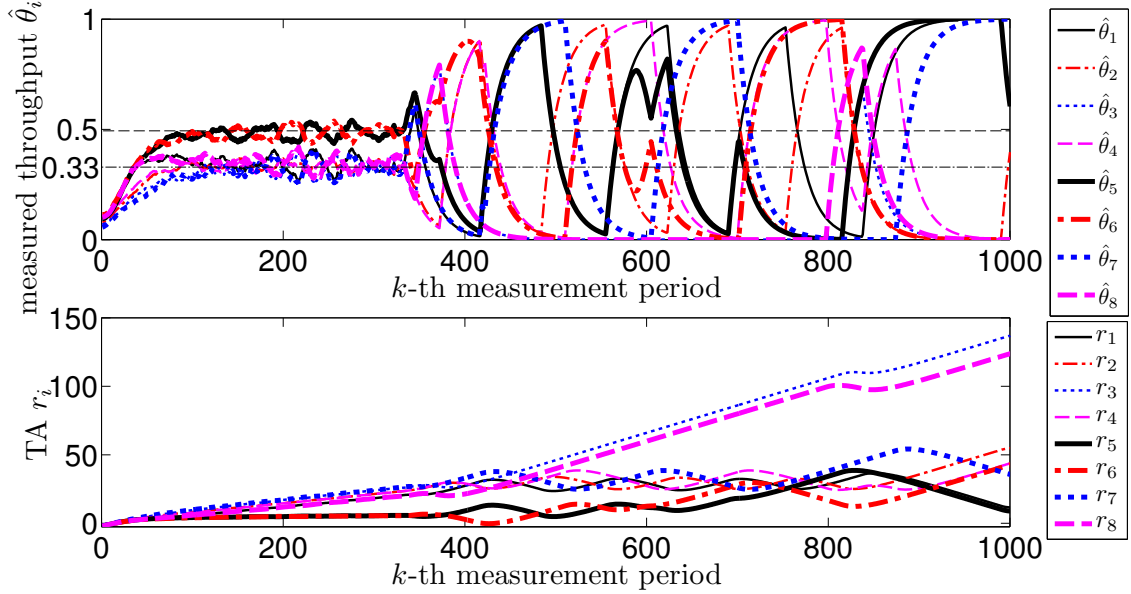} %,height=0.5\linewidth
\caption{Unstable Network Behavior without Price Control from BS} \label{unstable}
%\vspace{-1.5em}
\end{figure}

Finally, we illustrate the outcome of the CSMA game when there is no price control from the BS and when the collective target rates are outside the feasible throughput region $\bar{\mathcal{C}}$ (corresponding to $r_{max}=+\infty$). Suppose the 8 users in Fig. \ref{asym8} all desire a target rate of 0.5 (which is larger than the upper bound 0.33 achieved by the Stackelberg game in Section \ref{oscilate}), and they choose their TAs as in (\ref{IterationDynamics}) in order to achieve their own target rates. The behaviors of the D2D links are plotted in Fig. \ref{unstable}. Since the target rates are not achievable even when $r_{max}=+\infty$, each selfish user chooses an ever-increasing TA when its throughput is below its own target rate. When the TAs are high enough, the users intermittently capture the whole channel and prevent their neighbors from transmitting for a long time, which results in unstable network behavoirs and large access delay.
On the other hand, with the pricing mechanism, the network is stable and the maximum total throughput $g(M^{opt})$ is achieved as in Fig. \ref{Mrmax}, according to the $r_{max}$ value chosen. Therefore, the BS plays an important role in tuning the service price so that the network always operates within the feasible throughput region.

\section{Conclusions and Future Work}\label{Summary}
We study a group of D2D links which shares a dedicated inband overlay channel via CSMA. The ICN model is leveraged on to analyze their behaviors and interactions under spatial reuse. We further assume that the D2D links have heterogeneous rate requirements and different willingness to pay, and they act non-altruistically to achieve their target rates and maximize their own payoffs. To manage such non-cooperative user dynamics, we propose a Stackelberg game in which the BS in the cellular network acts as a Stackelberg leader to regulate the D2D link transmissions by modifying the service price, so that the total throughput is maximized while the heterogeneous target rates of the D2D links can all be satisfied. The problem is shown to be quasi-convex and can be solved by a sequence of equivalent convex optimization problems. The pricing strategies are designed so that the network always operates within the capacity region. The results are verified by simulations.
The joint optimization of D2D link scheduling and cellular data off-loading is our future work.

\bibliography{IEEEabrv,AlohaGames}

\end{document}